\documentclass[pdflatex,sn-mathphys-ay, iicol]{sn-jnl}

\usepackage{graphicx}%
\usepackage{multirow}%
\usepackage{amsmath,amssymb,amsfonts}%
\usepackage{amsthm}%
\usepackage{mathrsfs}%
\usepackage[title]{appendix}%
\usepackage{xcolor}%
\usepackage{textcomp}%
\usepackage{manyfoot}%
\usepackage{booktabs}%
\usepackage{algorithm}%
\usepackage{algorithmicx}%
\usepackage{algpseudocode}%
\usepackage{listings}%
\usepackage{comment}
\usepackage{bm, bbm}
\usepackage{makecell}

\theoremstyle{thmstyleone}%
\newtheorem{theorem}{Theorem}
\newtheorem{proposition}[theorem]{Proposition}%

\theoremstyle{thmstyletwo}%
\newtheorem{remark}{Remark}%

\theoremstyle{thmstylethree}%
\newtheorem{definition}{Definition}%

\newcommand{\Pb}{\mathbb{P}}

\begin{document}

\title[]{Parameter Estimation for Time-Scaled Inhomogeneous Phase-Type Distributions from Discrete Observations}


\author[1]{\fnm{Fernando} \sur{Baltazar-Larios}}\email{fernandobaltazar@ciencias.unam.mx}
\equalcont{Both authors contributed equally to this work.}

\author*[2]{\fnm{Alejandra} \sur{Quintos}}\email{alejandra.quintos@wisc.edu}
\equalcont{Both authors contributed equally to this work.}

\affil[1]{\orgdiv{Facultad de Ciencias}, \orgname{Universidad Nacional Aut\'onoma de M\'exico}, \orgaddress{\city{CdMx}, \country{M\'exico}, \postcode{A.P. 20-726, 01000}}}

\affil[2]{\orgdiv{Department of Statistics}, \orgname{University of Wisconsin}, \orgaddress{\city{Madison}, \state{Wisconsin}, \country{USA}, \postcode{53706}}}


\abstract{Inhomogeneous phase-type (IPH) distributions extend classical phase-type (PH) models by allowing transition intensities to vary over time, offering greater flexibility for modeling heavy-tailed distributions or time-dependent absorption phenomena. Statistical inference for these models has largely assumed that absorption times, or entire trajectories, are observed exactly. In many applications, however, the process is observed only at discrete, irregularly spaced time points, so that transition and absorption times are unknown and estimation becomes a missing-data problem. We address this setting for the subclass with time-scaled sub-intensity matrices $\boldsymbol{\Lambda}(t) = h_{\beta}(t)\boldsymbol{\Lambda}$, which admits a time transformation to a homogeneous Markov jump process (MJP). We develop an inference framework that combines Markov-bridge data augmentation with a Stochastic Expectation-Maximization (SEM) algorithm: at each iteration the latent continuous-time trajectories are simulated conditionally on the discrete observations, and the parameters are then updated by maximizing the resulting complete-data likelihood. The baseline sub-intensity matrix $\bm \Lambda$ is updated by its closed-form complete-data maximum-likelihood estimator, while the time-scaling parameter $\beta$ is refined by gradient ascent on the same complete-data log-likelihood. The reported estimators are thus obtained from complete-data maximum-likelihood updates, avoiding constrained nonlinear optimization of the observed-data likelihood. Through simulation studies for the matrix-Gompertz and matrix-Weibull families, and a real-data application to coronary allograft vasculopathy (CAV) progression, we demonstrate that the proposed approach provides an accurate and computationally tractable tool for fitting time-scaled IPH models to irregular multi-state data.}




\maketitle

\section{Introduction}\label{sec1}

Phase-type (PH) distributions describe the time until absorption of a finite-state homogeneous Markov jump process (MJP) and form a flexible class of distributions with rich structural and analytical properties \citep{Neus_75}. Inhomogeneous phase-type (IPH) distributions, introduced in \cite{Alb-Bla-19}, extend this framework by defining the absorption time of a finite-state inhomogeneous Markov jump process (IMJP), where transition intensities vary over time. This added flexibility enables IPH distributions to capture phenomena such as heavy-tailed behaviour, non-constant hazard rates, and time-varying transition dynamics that cannot be represented within the homogeneous PH family. Many classical heavy-tailed distributions (including Pareto, Weibull, and generalized extreme value) can be expressed as the distribution of a time-transformed exponential random variable, situating them naturally within the IPH framework.

IPH models have seen increasing use across a wide range of applied domains, including survival analysis \citep{Bao-11}, queueing systems \citep{zwart2001queueing}, finance \citep{bradley2003}, insurance \citep{Ahmad.Bladt.2023}, and epidemiology \citep{barraza2020non}. Despite their flexibility, statistical inference for general IPH distributions is challenging: the time-dependent infinitesimal generator rarely admits closed-form transition probabilities, and the intensity matrices at different time points typically do not commute. These issues complicate the likelihood evaluation and make parameter estimation difficult.

A notable subclass of IPH distributions alleviates these challenges by assuming that the sub-intensity matrix can be written in the time-scaled form
$$\bm \Lambda(t) = h_{\beta}(t)\bm \Lambda,$$
where $h_{\beta}$ is a nonnegative scaling function. This structure ensures that the sub-intensity matrices commute and that the underlying inhomogeneous process can be mapped to a homogeneous Markov jump process via a time transformation. The resulting tractability has motivated several recent estimation procedures. For example, \cite{Al-bla-ys-20} propose an expectation-maximization (EM) algorithm for datasets where absorption times are fully observed, and \cite{Ahm-bla-bla-24} study piecewise-constant models for the time-dependent sub-intensity matrix.

In many practical settings, however, the process is not observed continuously: only the states at discrete and irregular time points are available, and the exact transition times between states remain unobserved. This situation naturally arises in multi-state panel data, such as the progression of chronic diseases monitored at periodic clinical visits. In such cases, estimation becomes a missing-data problem in which both transition times and absorption times must be reconstructed or inferred.

In this work, we develop a statistical inference framework for time-scaled IPH distributions $\bm \Lambda(t) = h_{\beta}(t)\bm \Lambda$ when the underlying IMJP is observed only at discrete time points. Our method combines Markov-bridge simulation with a Stochastic Expectation-Maximization (SEM) algorithm to reconstruct latent continuous-time trajectories and estimate the baseline sub-intensity matrix $\bm \Lambda$ and the time-scaling parameter $\beta$. Conditional on the simulated complete paths, the update for $\bm \Lambda$ is available in closed form as the complete-data MLE, while $\beta$ is refined through a gradient-based maximization of the complete-data log-likelihood. Every iteration therefore returns the maximum-likelihood estimators associated with a simulated complete-data realization, and the reported estimates are obtained from such complete-data MLE updates rather than from direct optimization of the observed-data likelihood. This yields a flexible and computationally efficient likelihood-based estimation procedure applicable to irregularly spaced observations. Furthermore, we establish the theoretical conditions under which the baseline sub-intensity matrix and the time-scaling function are identifiable, thereby ruling out parameter confounding.

We assess the performance of the proposed method through simulation studies for the matrix-Gompertz and matrix-Weibull families, demonstrating accurate parameter recovery across a range of settings. We then apply the method to coronary allograft vasculopathy (CAV) data, a motivating example involving irregular multi-state observations in a medical setting. We further compare the fitted inhomogeneous model with its homogeneous counterpart, highlighting the importance of accounting for time-varying transition intensities in real-world applications.

The remainder of the paper is organized as follows. Section~\ref{ssec:def} reviews IPH distributions, their main properties, and the subclass considered in this work. Section~\ref{sect.Statistical.Inference} presents the proposed estimation procedure in detail. The simulation results are given in Section~\ref{Sect.Simulation.Study}, and Section~\ref{sec:real} contains the application to the CAV dataset and the comparison with a homogeneous PH model.

\section{Inhomogeneous Phase-Type Distributions}\label{ssec:def} 
Let $\bm X= \left\{X_t \right\}_{0\leq t \leq T}$ be an IMJP taking values in a finite state space $E=\{1,2,\ldots,n,n+1\}$, where the states $1, \ldots, n$ are transient, and the state $n+1$ is the unique absorbing state. This process is defined by the initial distribution $\bm \alpha=(\bm \pi,0)$, where $\bm \pi=(\pi_1,\ldots,\pi_n)$, and by the infinitesimal generator $ \bm Q(t)$, given by
$$ \bm Q(t):=\begin{pmatrix}
	\bm \Lambda(t)& \bm \lambda(t)\\
	\bm 0_n & 0 
\end{pmatrix},$$
where $\bm \Lambda(t)$ is an $n \times n$ sub-intensity matrix whose entries depend on time, $ \bm \lambda(t) = - \bm \Lambda (t) \mathbbm{1}_n$ is the exit-rate vector (with $\mathbbm{1}_n$ denoting an $n$-dimensional column vector of ones),  and $\bm 0_n$ is the $n$-dimensional zero row vector.

\begin{definition}
	We define the time until absorption of $\bm X$ as
	$$\tau:=\inf\left\{t>0 \mid X_t=n+1\right\},$$
	and say that $\tau$ follows an IPH distribution with parameters $(\bm \pi,\bm \Lambda(t))$, denoted as $\tau \sim \mathrm{IPH}(\bm \pi, \bm \Lambda(t))$.
\end{definition}

Let $\bm P(s,t) = \left\{p_{xy}(s,t)\right\}_{x,y\in E}$ ($0 < s < t$) denote the transition probability matrix of the process $\bm X$, where
$$p_{xy} (s,t)=\Pb(X_t=y|X_s=x),$$
which can be expressed in terms of the product integral of the intensity matrix as \citep{Sla}
$$\bm P(s,t)=\prod_s^t(\bm I+\bm Q(u)du),$$
where the product integral is defined as
\begin{multline*}
	\prod_s^t(\bm I+\bm Q(u)du):= \bm I + \\
	\sum_{k=1}^{\infty}\int_s^t\int_s^{u_k}\cdots\int_s^{u_2}\bm Q(u_1)\cdots\bm Q(u_k)du_1\cdots du_k.
\end{multline*}

The density and distribution functions of $\tau$ \citep{Alb-Bla-19} are given by
\begin{equation}\label{den_iph}
	f_{\tau}(x)=\bm {\pi}\prod_0^x(\bm I+\bm\Lambda(u)du) \bm \lambda(x), 
\end{equation}
and 
\begin{equation}\label{dis_iph}
	F_{\tau}(x)=1-\bm {\pi}\prod_0^x(\bm I+\bm\Lambda(u)du)\mathbbm{1}_n.
\end{equation}

If the matrices $\bm \Lambda(t)$ commute for all $t$, then the density \eqref{den_iph} simplifies to
$$f_{\tau}(x)=\bm {\pi}\mbox{exp}\left(\int_0^x\bm\Lambda(t)dt\right)\bm\lambda(x),    
$$
and the distribution function \eqref{dis_iph} becomes 
$$F_{\tau}(x)=1-\bm {\pi}\mbox{exp}\left(\int_0^x\bm\Lambda(t)dt\right)\mathbbm{1}_n.   
$$

In particular, we consider the subclass of IPH distributions in which the sub-intensity matrix takes the form 
\[
\bm \Lambda(t) = h_{\beta}(t)\bm \Lambda,
\]
where $h_{\beta}(t)$ is a known non-negative real-valued function depending on a parameter $\beta$, and $\bm \Lambda$ is a non-time-varying $n\times n$ sub-intensity matrix. We denote this subclass as 
\[
\tau \sim \mathrm{IPH}(\bm \pi, \bm \Lambda, h_{\beta}).
\]  

In this case, all matrices commute and there exists a function $g_{\beta}$ defined through its inverse function as
$$g_{\beta}^{-1}(t)=\int_0^th_{\beta}(s)ds,$$
such that 
\[
\tau \overset{d}{=} g_{\beta}(\rho), \quad \text{where } \rho \sim \mathrm{PH}(\bm \pi, \bm \Lambda).
\]

If  $\tau\sim\mbox{IPH}(\bm \pi,\bm \Lambda,h_{\beta})$, then the density function $f_\tau$ is given by 
\begin{equation}\label{den_sca}
	f_\tau(x;\bm \pi, \bm \Lambda,\beta)=h_{\beta}(x)\bm {\pi}\mbox{exp}\left(\int_0^x h_{\beta}(t)dt\bm\Lambda\right)\bm\lambda, 
\end{equation}
and the distribution function $F_\tau$ is given by 
\begin{equation}\label{dis_sca}
	F_\tau(x; \bm \pi, \bm \Lambda,\beta)=1-\bm {\pi}\mbox{exp}\left(\int_0^x h_{\beta}(t)dt\bm\Lambda\right)\mathbbm{1}_n.
\end{equation}
Moreover, the expected value of $\tau$ \citep{Alb-Bla-19}  is 
\begin{equation}\label{esp_sca}
	\mathbb{E}[\tau]=\bm {\pi}L_g(-\bm\Lambda)\bm \lambda,
\end{equation}
where $L_g(s) $ is the Laplace transform\footnote{$L_g(s)=\int_0^{\infty}g(t)e^{-st}dt$.}, which exists for all $s > \max_i Re(\gamma_i)$ (whenever the Laplace transform exists), with $\gamma_i$ denoting the $i$-th eigenvalue of $\bm \Lambda$.

We denote by $\bm Y=\left\{Y_t \right\}_{0\leq t \leq T}$ the homogeneous MJP with infinitesimal generator 
$$\bm Q=\begin{pmatrix}
	\bm \Lambda& \bm \lambda\\
	\bm 0_n & 0 
\end{pmatrix},$$
where $\bm \Lambda=\{\lambda_{xy}\}_{x,y=1}^n$ is an $n\times n$ sub-intensity matrix and $\bm \lambda=(\lambda_1,\ldots,\lambda_n)^T$ is a column vector of dimension $n$ such that $ \bm \lambda = - \bm \Lambda \mathbbm{1}_n$ with $\lambda_x$ denoting the exit rate from state $x=1,\ldots,n$. This process shares the same structural parameters $(\bm \pi, \bm \Lambda)$ as the IMJP $\bm X$, with the key difference that in $\bm X$, the sub-intensity matrix 
is scaled by the time-dependent function $h_\beta(t)$. 

\section{Statistical Framework for Estimation} \label{sect.Statistical.Inference}
\subsection{Observation Scheme and Data} \label{subsect:observation.scheme}
Consider $K \in \mathbb{N}$ independently and identically distributed (i.i.d.) paths of an underlying IMJP $\bm{X}$, associated with the IPH distribution $\tau \sim \mathrm{IPH}\left(\bm{\pi}, \bm{\Lambda}, h_\beta\right)$, defined over the time interval $[0, T]$, where $T > 0$. These paths are not continuously observed over time. Instead, for each path $k = 1, \ldots, K$, we observe it only at a discrete set of time points denoted by  
\begin{equation*}
	\boldsymbol{T}^{(k)} := \left\{ 0 = t_{k,0} < t_{k,1} < \cdots < t_{k,m_k} \leq T \right\},
\end{equation*}
where $m_k \in \mathbb{N}$ indicates that the $k$-th path is observed at $m_k + 1$ time points, from $t_{k,0} = 0$ to $t_{k,m_k} \leq T$.

Let $\bm{X}^d$ denote the observed discrete-time trajectories:
\begin{multline*}
	\bm{X}^d := \left\{ X_{t_{k, 0}} = x_{t_{k,0}},\, X_{t_{k, 1}} = x_{t_{k,1}},\, \dots,\, \right.\\
	\left. X_{t_{k, m_k}} = x_{t_{k,m_k}} \right\}_{k=1}^K,
\end{multline*}
where each $x_{t_{k,j}} \in E$. For instance, $x_{t_{k, 0}}$ is the initial state of the $k$-th path, and $x_{t_{k, m_k}}$ is its last observed state.

\begin{remark}\label{rmk:incomplete.censored.data}
    Each path is observed only at the discrete time points in $\boldsymbol{T}^{(k)}$, and only until absorption or censoring; we therefore allow $t_{k,m_k} \leq T$, so that not all trajectories are followed to the terminal time $T$. The absorbing state, if observed at all, can only appear at the final observation time: $X_{t_{k,j}} \neq n+1$ for all $j < m_k$, while $X_{t_{k,m_k}}$ may or may not equal $n+1$. Consequently, neither the transition times between transient states nor the absorption time are available. In particular, if $X_{t_{k,j-1}}$ is transient and $X_{t_{k,j}} = n+1$, absorption occurred at some unknown instant in $\left(t_{k,j-1}, t_{k,j}\right]$. Our methodology treats these latent transition and absorption times as missing data, simulated as part of the estimation procedure.
\end{remark}

Our objective is to estimate the parameters $\left(\bm{\pi}, \bm{\Lambda}, \beta\right)$\footnote{We assume that the functional form of $h_\beta$ is known up to the parameter $\beta$.}. In what follows, we describe the estimation procedure in detail. A summarized version is presented at the end of this section as Algorithm~\ref{main.algorithm}.

\begin{remark} \label{rmk.paper.mogens}
	In \cite{Al-bla-ys-20}, the parameters $\left(\bm{\pi}, \bm{\Lambda}, \beta\right)$ were estimated using the EM algorithm under the assumption that the data consist of absorption times observed explicitly. In contrast, our setting considers discretely observed trajectories, where absorption times are generally unobserved and must be inferred indirectly.
\end{remark}

Since the initial state of each of the $K$ paths is known, estimation of $\bm{\pi}$ is straightforward via the empirical distribution. Specifically, let $\hat{\pi}_x$ denote the proportion of paths that start in state $x \in \left\{1, 2, \dots, n \right\}$, i.e.,
\begin{equation}
	\hat{\pi}_x := \frac{1}{K} \sum_{k=1}^K \mathbb{I}\left\{ x_{t_{k,0}} = x \right\},
\end{equation}
then the estimator is given by
\begin{equation}\label{estimation.pi}
	\bm{\hat{\pi}} := \left( \hat{\pi}_1, \hat{\pi}_2, \dots, \hat{\pi}_n \right).
\end{equation}

Because the initial states are observed exactly, $\bm {\hat{\pi}}$ involves no missing data and is unaffected by the reconstruction of the latent trajectories. It is therefore computed once, prior to the iterative procedure, and held fixed.

Estimation of $\bm{\Lambda}$ and $\beta$ is performed iteratively and simultaneously by applying a time transformation that converts each inhomogeneous path into an equivalent homogeneous one. Specifically, if $t$ represents a time point on the inhomogeneous scale, then as shown, for example, in \cite{Hubbard.et.al}, the corresponding homogeneous time $s$ is given by
\begin{equation} \label{eq:Connection.Inhomogeneous.Homogeneous}
	s = g^{-1}_{\beta}(t) \ = \int_0^t h_\beta(s)ds.
\end{equation}

Define the time-transformed discrete observations
\begin{multline*}
	\bm{Y}^d := \left\{ Y_{s_{k,0}} = y_{s_{k,0}} = x_{t_{k,0}},\, Y_{s_{k,1}} = y_{s_{k,1}} = \right. \\
	\left. x_{t_{k,1}},\, \dots,\, Y_{s_{k,m_k}} = y_{s_{k,m_k}} = x_{t_{k,m_k}} \right\}_{k=1}^K,
\end{multline*}
where $s_{k,j} = g^{-1}_{\beta}(t_{k,j})$ for $j = 0, \dots, m_k$ and $k = 1, \dots, K$. Here, $\bm{Y}^d$ represents the discrete observations of a homogeneous MJP $\bm{Y}$ associated with the IMJP $\bm{X}$. Furthermore, the associated distribution of $\bm{Y}$ is the standard phase-type distribution $\rho \sim \mathrm{PH}(\bm{\pi}, \bm{\Lambda})$.

\subsection{Methodological Preliminaries}
For the time-scaled subclass considered here, the observed-data likelihood can in principle be written explicitly, since the transition probabilities between observation times are obtained from matrix exponentials after the time transformation $s = g_{\beta}^{-1}(t)$. Thus, direct maximum-likelihood estimation is possible. However, this approach requires constrained nonlinear optimization over the entries of $\bm \Lambda$  and $\beta$, involving repeated evaluations of matrix exponentials and, in gradient-based implementations, derivatives of these matrix exponentials with respect to both $\bm \Lambda$ and $\beta$. We therefore adopt a stochastic complete-data approach: by augmenting the missing continuous-time trajectories through Markov bridges, the observed-data optimization problem is replaced by a sequence of complete-data maximum-likelihood problems, each of which is either available in closed form or reduces to a one-dimensional maximization. This yields a computationally simple and structurally interpretable likelihood-based procedure for irregularly observed trajectories.

Before presenting our estimation algorithm, we review two simplified estimation scenarios that have been studied in the literature, each assuming that one of the parameters is known and that the underlying process is observed continuously. In these idealized settings, all transition times and visited states are observed exactly. Although our actual data are only observed at discrete time points, these cases serve as building blocks for the more realistic problem of jointly estimating $\bm{\Lambda}$ and $\beta$ from discretely observed paths. The two cases are:

\begin{itemize}
	\item $\beta$ is known and only $\bm{\Lambda}$ needs to be estimated \citep{bill}.
	\item $\bm{\Lambda}$ is known and only $\beta$ needs to be estimated, either by analytically deriving the MLE or by using numerical optimization methods \citep{bottou2018optimization}.
\end{itemize}

For both cases, assume that we have $K \in \mathbb{N}$ i.i.d. sample paths of an underlying IMJP $\bm{X}$ associated with the IPH distribution $ \tau \sim \mathrm{IPH}\left(\bm{\pi}, \bm{\Lambda}, h_\beta\right) $, defined on the time interval $ [0, T] $, where $ T > 0 $. As mentioned previously, in contrast to our primary setting involving discrete observations $\boldsymbol{X}^d$, we momentarily assume that these paths are observed continuously in time. That is, for each path $k = 1, \ldots, K$, we observe the exact times of all state transitions and the specific states visited. We denote this collection of continuously observed paths by $\bm{X}^c$, which can then be written as:
\begin{multline} \label{Continuous.Data.IPH}
	\bm{X}^c := \left\{ X_{\tilde{t}_{k, 0}} = x_{\tilde{t}_{k,0}},\, X_{\tilde{t}_{k, 1}} = x_{\tilde{t}_{k,1}},\, \dots,\, \right. \\
	\left. X_{\tilde{t}_{k, n_k}} = x_{\tilde{t}_{k,n_k}} \right\}_{k=1}^K,
\end{multline}
where $x_{\tilde{t}_{k,j}} \in E$ denotes the state entered at transition time $\tilde{t}_{k,j}$ for $j = 0, \dots, n_k$. For example, $x_{\tilde{t}_{k,0}}$ represents the initial state of the $k$-th path, $\tilde{t}_{k,1}$ is the time of the first transition to state $x_{\tilde{t}_{k,1}}$, and $\tilde{t}_{k, n_k}$ is the final observed transition time into state $x_{\tilde{t}_{k, n_k}}$. Note that for each $k = 1, \ldots, K$, the final transition time satisfies $\tilde{t}_{k, n_k} \leq T$.
\begin{remark} \label{rmk.discrete.vs.continuous.times}
	Note that $m_k$ and $n_k$ are not necessarily equal. The quantity $m_k + 1$ represents the number of discrete observation time points for the $k$-th path, whereas $n_k$ denotes the number of actual state transitions under continuous observation. Additionally, the transition times $\tilde{t}_{k,j}$ in the continuous setting generally do not coincide with the discrete observation times $t_{k,j}$.
\end{remark}
Based on this continuously observed data $\bm{X}^c$, we can explicitly define the complete-data likelihood function on the original inhomogeneous scale. The likelihood of a continuous-time Markov trajectory is constructed by multiplying the initial state probabilities, the instantaneous transition intensities evaluated exactly at the jump times, and the survival probabilities of remaining in the visited states during the inter-jump intervals (see \cite{Ander}). For our time-scaled IPH model, the transition intensity from state $x$ to $y$ at time $t$ is $h_\beta(t)\lambda_{xy}$. Consequently, the holding rate in state $x$ is $h_\beta(t)\left(\sum_{y \neq x} \lambda_{xy} + \lambda_x\right)$, and the probability of no transitions occurring out of state $x$ during an interval $[t_1, t_2]$ is given by $\exp\left(-\left(\sum_{y \neq x} \lambda_{xy} + \lambda_x\right) \int_{t_1}^{t_2} h_\beta(u) du\right)$.

Let $B_x$ be the number of paths starting in state $x$, $N_{xy}(T)$ be the total number of observed transitions from state $x$ to $y$ across all paths, $N_x(T)$ be the total number of transitions from state $x$ to the absorbing state, and $R_x^\beta(T) = \int_0^T \mathbbm{1}_{\left\{X_t = x\right\}} h_\beta(t) dt$ be the scaled occupation time in state $x$. By aggregating the transition and survival components across all $K$ paths, the complete-data likelihood function $L^c_T(\bm{\Lambda}, \beta)$ is given by:
\begin{multline} \label{eq.Likelihood.Complete.Inhomogeneous}
    L^c_T(\bm\Lambda, \beta) = \left( \prod_{x=1}^n \pi_x^{B_x} \right) \left( \prod_{k=1}^K \prod_{j=1}^{n_k} h_\beta\left(\tilde{t}_{k,j}\right) \right) \\
    \left( \prod_{x=1}^n \prod_{\substack{y=1 \\ y \neq x}}^n \lambda_{xy}^{N_{xy}(T)} \right) \left( \prod_{x=1}^n \lambda_x^{N_x(T)} \right) \\
    \exp\left( -\sum_{x=1}^n \left(\sum_{\substack{y=1 \\ y \neq x}}^n \lambda_{xy} + \lambda_x \right) R_x^\beta(T) \right).
\end{multline}
Taking the natural logarithm yields the complete-data log-likelihood function:
\begin{multline} \label{eq.LogLikelihood.Complete.Inhomogeneous}
    \ell^c_T(\bm{\Lambda}, \beta) = \sum_{x=1}^n B_x \log\left(\pi_x\right) + \sum_{k=1}^K \sum_{j=1}^{n_k} \log h_\beta\left(\tilde{t}_{k,j}\right) \\
    + \sum_{x=1}^n \sum_{\substack{y=1 \\ y \neq x}}^n N_{xy}(T) \log\left(\lambda_{xy}\right) \
    + \sum_{x=1}^n N_x(T) \log\left(\lambda_x\right) \\
    - \sum_{x=1}^n \left(\sum_{\substack{y=1 \\ y \neq x}}^n \lambda_{xy} + \lambda_x \right) R_x^\beta(T).
\end{multline}
This complete-data log-likelihood serves as the foundational objective function for estimating the parameters when continuous paths are available. Note that its first term depends only on the observed initial-state counts $B_x$ and is constant with respect to $\bm \Lambda$ and $\beta$. Hence, it plays no role in the iterative procedure.

In Section \ref{subsect:handling.discrete.times}, we describe how to handle the more realistic setting in which paths are only observed at discrete time points. This requires using Markov bridges to simulate continuous-time trajectories that are consistent with discretely observed data, assuming the model parameters are known.

Building on this, in Section \ref{subsect:em-like.estimation}, we present a unified perspective that connects all three components (estimation under known $\beta$, estimation under known $\bm{\Lambda}$, and handling discretely observed data) under a common EM-like framework. This conceptual unification will serve as a guiding principle for the full estimation procedure presented in Section ~\ref{subsection.full.estimation.procedure}.

Finally, in Section \ref{subsect:identifiability}, we address the theoretical requirement of parameter identifiability, establishing the conditions under which the baseline sub-intensity matrix and the time-scaling function can be uniquely determined without confounding.

\subsubsection{\texorpdfstring{$\beta$}{Beta} is known} \label{subsect:beta.is.known}

We first consider the case where the parameter $\beta$ is known, and we want to estimate $\bm{\Lambda}$ from a continuously observed set of paths $\bm{X}^c$ on the interval $[0,T]$. Returning to the complete-data log-likelihood on the original inhomogeneous scale defined in Equation \eqref{eq.LogLikelihood.Complete.Inhomogeneous}, when $\beta$ is fixed, the term $\sum_{k=1}^K \sum_{j=1}^{n_k} \log h_\beta\left(\tilde{t}_{k,j}\right)$ is constant with respect to $\bm{\Lambda}$.

To maximize the remaining terms with respect to $\bm{\Lambda}$, instead of working directly with the inhomogeneous paths $\bm{X}^c$, we use the time transformation $g_\beta^{-1}$ defined in equation \eqref{eq:Connection.Inhomogeneous.Homogeneous} to transform all observed inhomogeneous time points in \eqref{Continuous.Data.IPH} to their homogeneous equivalents. Let $\bm{Y}$ be the corresponding homogeneous MJP defined on the transformed time interval $\left[0, S\right]$, where $S = g_\beta^{-1}(T)$. The mapped paths, denoted by $\bm{Y}^c$, are given by:
\begin{multline*}
	\bm{Y}^c := \left\{ Y_{\tilde{s}_{k,0}} = y_{\tilde{s}_{k,0}} = x_{\tilde{t}_{k,0}},\, Y_{\tilde{s}_{k,1}} = y_{\tilde{s}_{k,1}} = \right. \\
	\left. x_{\tilde{t}_{k,1}},\, \dots,\, Y_{\tilde{s}_{k,n_k}} = y_{\tilde{s}_{k,n_k}} = x_{\tilde{t}_{k,n_k}} \right\}_{k=1}^K,
\end{multline*}
where $\tilde{s}_{k,j} = g^{-1}_{\beta}(\tilde{t}_{k,j})$ for $j = 0, \dots, n_k$ and $k = 1, \dots, K$.

Because the topological properties of the paths are invariant under this strictly increasing mapping, the sequence of visited states remains the same, and the transition counts are preserved. That is, $N_{xy}(T) = N_{xy}(S)$ and $N_x(T) = N_x(S)$.

Furthermore, the amount of time the mapped paths $\bm{Y}^c$ spend in state $x$ during the interval $[0,S]$ is exactly the scaled occupation time of the original paths $\bm{X}^c$ on $[0,T]$. Specifically, by applying the change of variables $ds = h_\beta(t)dt$, we obtain:
\begin{multline} \label{eq.Rx.equivalence}
    R_x^\beta(T) = \int_0^T \mathbbm{1}_{\left\{X_t = x\right\}} h_\beta(t) dt \\
    = \int_0^S \mathbbm{1}_{\left\{Y_s=x\right\}} ds = R_x(S).
\end{multline}

Consequently, substituting these equivalences back into Equation \eqref{eq.LogLikelihood.Complete.Inhomogeneous} and dropping the constant $\beta$ term demonstrates that maximizing the complete-data log-likelihood of the inhomogeneous model with respect to $\bm{\Lambda}$ is exactly equivalent to maximizing the complete-data log-likelihood function of the homogeneous Markov jump process $\bm{Y}$ evaluated on the mapped continuous paths $\bm{Y}^c$ over $[0, S]$. This homogeneous log-likelihood $l_S^c(\bm{\Lambda})$ is written as (see, e.g., \cite{bill}):
\begin{multline}\label{loglike}
		\ell^c_S(\bm{\Lambda}) 
		= \sum_{x=1}^{n} B_x \log(\pi_x) + \\
		 \sum_{x=1}^{n} \sum_{\substack{y=1 \\ y \neq x}}^{n} N_{xy}(S) \log(\lambda_{xy}) 
		- \sum_{x=1}^{n} \sum_{\substack{y=1 \\ y \neq x}}^{n} \lambda_{xy} R_x(S)
		+ \\
		\sum_{x=1}^{n} N_x(S) \log(\lambda_x)
		- \sum_{x=1}^{n} \lambda_x R_x(S).
\end{multline}

Maximizing this log-likelihood yields the MLEs for $ \bm{\pi} $ and $ \bm{\Lambda} $, which are given by:
\begin{align*}
	\hat{\pi}_x = \frac{B_x}{K}, & \qquad
	\hat{\lambda}_{xy}(S) = \frac{N_{xy}(S)}{R_x(S)}, \\
	 \hat{\lambda}_x(S) & = \frac{N_x(S)}{R_x(S)},
\end{align*}
provided that \( R_x(S) > 0 \) for all \( x \in \{1, \ldots, n\} \).

We summarize this result in the following proposition.

\begin{proposition}[Maximum Likelihood Estimators in the Homogeneous Timeline]\label{prop:MLEs}
	Suppose $\beta$ is known and that the continuous-time sample paths $\bm{X}^c$ of the time-inhomogeneous Markov jump process $\bm{X}$ are fully observed over the interval $[0, T]$. Let $\bm{Y}^c$ denote the continuous-time paths obtained by mapping $\bm{X}^c$ via the time transformation $s = g^{-1}_{\beta}(t)$. Then, the MLEs for the initial distribution $\bm{\pi}$ and the sub-intensity matrix $\bm{\Lambda}$, based on the transformed paths $\bm{Y}^c$ observed over the homogeneous time interval $[0, S]$, where $S := g^{-1}_{\beta}(T)$, are given by:
	\begin{align} \label{mles}
		\hat{\pi}_x = \frac{B_x}{K}, & \qquad 
		\hat{\lambda}_{xy}(S) = \frac{N_{xy}(S)}{R_x(S)}, \nonumber \\
        \hat{\lambda}_x(S) & =  \frac{N_x(S)}{R_x(S)},
	\end{align}
	for all $x, y \in \{1, \dots, n\}$ with $x \ne y$, provided that $R_x(S) > 0$ for all $x$.
\end{proposition}

\begin{remark}
	The diagonal elements of the sub-intensity matrix, \( \lambda_{xx} \), are not estimated directly. Instead, they are obtained using the fact that each row of \( \bm{\Lambda} \) must sum to zero. Specifically, for each transient state \( x \in \{1, \dots, n\} \),
	\begin{equation*}
		\hat{\lambda}_{xx}(S) = -\left( \sum_{\substack{y=1 \\ y \ne x}}^{n} \hat{\lambda}_{xy}(S) + \hat{\lambda}_x(S) \right).
	\end{equation*}
	This ensures that \( \bm{\Lambda} \) remains a valid sub-intensity matrix, where the diagonal elements represent the negative holding rates.
\end{remark}

\subsubsection{\texorpdfstring{$\bm \Lambda$}{Lambda} is known} \label{subsect:lambda.is.known}
In this subsection, we assume that the sub-intensity matrix $\bm{\Lambda}$ is known, and our goal is to estimate the parameter $\beta$ that maximizes the complete-data joint likelihood of $K$ i.i.d. paths of the IMJP observed continuously over the time interval $[0, T]$ (see equation \eqref{eq.Likelihood.Complete.Inhomogeneous}). When a closed-form expression for the MLE is not available, numerical optimization methods are commonly employed. In particular, we use the Gradient Ascent (GA) method \citep{bottou2018optimization}, an iterative optimization procedure that updates $\beta$ in the direction of the gradient.

Since $\bm{\pi}$, $\bm{\Lambda}$, and the functional form of $h_\beta$ are all assumed known (except for the parameter $\beta$), the complete-data log-likelihood function $\ell^c_T(\beta)$ associated with the continuously observed paths $X$ on $[0,T]$ depends on $\beta$ only through the jump times and the state occupation times. Dropping terms in equation~\eqref{eq.LogLikelihood.Complete.Inhomogeneous} that do not depend on $\beta$, the complete-data log-likelihood function is proportional to:
\begin{multline}\label{eq:log.lik.proportional.beta}
   \ell^c_T(\beta) \propto \sum_{k=1}^K \sum_{j=1}^{n_k} \log h_\beta\left( \tilde{t}_{k,j} \right) \\
    - \sum_{x=1}^n \left(\sum_{\substack{y=1 \\ y \neq x}}^n \lambda_{xy} + \lambda_x \right) R_x^\beta(T),
\end{multline}
where, as stated after equation~\eqref{Continuous.Data.IPH}, $\tilde{t}_{k,j}$ denotes the time of the $j$-th observed jump for the $k$-th path, $n_k$ is the total number of transitions for the $k$-th path, $\lambda_{xy}$ are elements of the $\boldsymbol{\Lambda}$ matrix, $\left( \lambda_1, \lambda_2, \dots, \lambda_n \right) = -\boldsymbol{\Lambda} \mathbbm{1}_n$, and $R_x^\beta\left(T\right) = \int_0^T \mathbbm{1}_{\left\{X_t = x\right\}} h_\beta(t) dt$ is the time spent in state $x$ scaled by the intensity function $h_\beta(t)$.

We estimate $\beta$ by maximizing the log-likelihood via the GA method. This requires evaluating the gradient of the complete-data log-likelihood function with respect to $\beta$, given by:
\begin{multline}
     \frac{\partial}{\partial \beta} \ell^c_T \left(\beta \right) = \sum_{k=1}^K \sum_{j=1}^{n_k} \frac{\frac{\partial}{\partial \beta} h_\beta\left(\tilde{t}_{k,j}\right)}{h_\beta\left(\tilde{t}_{k,j}\right)}   \\
     - \sum_{x=1}^n \left(\sum_{\substack{y=1 \\ y \neq x}}^n \lambda_{xy} + \lambda_x \right)\frac{\partial}{\partial \beta} R_x^\beta\left(T\right).
\end{multline}

\begin{remark}
    To justify differentiating under the integral sign for the interval term $\frac{\partial}{\partial \beta} R_x^\beta\left(T\right) = \frac{\partial}{\partial \beta} \int_0^T \mathbbm{1}_{\{X_t = x\}} h_\beta(t) dt$, we note that the limits of integration do not depend on $\beta$. We assume that $h_\beta(s)$ is continuously differentiable in $\beta$ for each $s \in [0, T]$, and that $\frac{\partial}{\partial \beta} h_\beta(s)$ is integrable over $[0, T]$. These regularity conditions justify the interchange via the Leibniz integral rule, yielding $\int_0^T \mathbbm{1}_{\{X_t = x\}} \frac{\partial}{\partial \beta} h_\beta(t) dt$.
\end{remark}

Given the gradient, the GA algorithm proceeds by iteratively updating $\beta$. Starting from an initial guess $\hat{\beta}^{(0)}$, each iteration updates the estimate according to:
\begin{multline}\label{eq:GD.update}
	\hat{\beta}^{(i+1)} = \max\left\{ \beta_{\min},\ \hat{\beta}^{(i)} + \right. \\
	\left. \eta \cdot \frac{\partial}{\partial \beta} \ell^c_T\left( \hat{\beta}^{(i)} \right)\right\},
\end{multline}
where $i \in \mathbb{N}$ denotes the iteration index, and $\eta > 0$ is a fixed learning rate (or step size). We fix the lower bound to $\beta_{\min} > 0$ to ensure that $\hat{\beta}$ remains strictly positive. In our implementation, we set $\beta_{\min} = 10^{-5}$, which is sufficiently small to avoid biasing the estimate while still preventing degenerate behavior.

The procedure is repeated until convergence, which, in our case, is defined as the first iteration $i$ such that:
\begin{equation} \label{eq:GD.convergence.criteria}
	\left| \ell^c_T\left(\hat{\beta}^{\left(i\right)}\right) - \ell^c_T\left(\hat{\beta}^{\left( i-1 \right)}\right) \right| < e_\ell,
\end{equation}
for a fixed convergence threshold $e_\ell > 0$.

\subsubsection{Handling Discrete Times} \label{subsect:handling.discrete.times}
To apply the continuous-time estimation methods described in sections \ref{subsect:beta.is.known} and \ref{subsect:lambda.is.known} to the discretely observed data $\boldsymbol{X}^d$, we must reconstruct plausible continuous-time trajectories of the underlying IMJP that are consistent with the observed states at the discrete time points.

Assume for now that both $\bm{\Lambda}$ and $\beta$ are known. The first step in this reconstruction is to transform the observed time points $\left( t_{k,1}, t_{k,2}, \dots, t_{k,m_k} \right)$ to the homogeneous time scale using the transformation $s_{k,j} = g^{-1}_{\beta}(t_{k,j})$, for $k = 1, \dots, K$ and $j = 1, \dots, m_k$. After simulating the corresponding continuous time trajectories in the homogeneous time scale, we map them back to the original inhomogeneous scale via the inverse transformation $\tilde{t}_{k,j} = g_{\beta}(\tilde{s}_{k,j})$, where each $\tilde{s}_{k,j}$ belongs to the set $\left\{ \tilde{s}_{k,1}, \tilde{s}_{k,2}, \dots, \tilde{s}_{k,n_k} \right\}$ denoting the simulated transition times of the homogeneous MJP.\footnote{We use the same notation convention discussed in Remark~\ref{rmk.discrete.vs.continuous.times}, distinguishing between observed discrete time points and latent transition times under continuous observation.}

A natural way to simulate such homogeneous MJP paths is to condition on the observed states at the transformed time points. This can be achieved using Markov bridges, which sample the latent transitions between consecutive observation times in the homogeneous time scale. These sampled paths yield reconstructed continuous-time trajectories that are consistent with the observed discrete-time states. We define Markov bridges formally below.

\begin{definition}\label{bridge_mjp}
	A Markov bridge associated with a homogeneous Markov jump process is the process $\{Y_t\}_{t \in [t_1, t_2]}$ conditioned on $Y_{t_1} = x$ and $Y_{t_2} = y$, where $t_1 < t_2$ and $x, y \in E$. We refer to such a bridge as a $(t_1, x, t_2, y)$-Markov bridge.
\end{definition}

Accordingly, we simulate sample paths from $(s_{k,j}, y_{s_{k,j}}, s_{k,j+1}, y_{s_{k,j+1}})$-Markov bridges for each $k = 1, 2, \dots, K$ and $j = 0, 1, \dots, m_k - 1$. Various methods for sampling Markov bridges have been extensively studied in the stochastic process and computational statistics literature. A summary and comparison of the most prominent approaches can be found in \cite{com}. In this work, we focus on the rejection sampling method, which is described in Algorithm~\ref{algorithm.markov.bridges}.

\begin{algorithm*}[htbp] 
	\caption{Sampling Markov Bridges via the Rejection Method}
	\label{algorithm.markov.bridges}
	\textbf{Input:} $\bm{X}^d = \left\{ x_{t_{k,0}},\, x_{t_{k,1}},\, \dots,\, x_{t_{k,m_k}} \right\}_{k=1}^K$, $\bm{\Lambda}$, and $\beta$.\\
	\textbf{Output:} $K$ inhomogeneous, continuous-time paths $\left\{ x_{\tilde{t}_{k,0}},\, x_{\tilde{t}_{k,1}},\, \dots,\, x_{\tilde{t}_{k,n_k}} \right\}_{k=1}^K$.
	
	\begin{algorithmic}[1]
		\For{$k \in \left\{1, 2, \dots, K \right\}$}
			\State \label{algorithm.markov.bridges.transform.to.homogeneous}Transform observation times to the homogeneous scale: $s_{k,j} = g^{-1}_{\beta}(t_{k,j})$ for $j = 0, \dots, m_k$.
			\State Define $\bm{Y}^d_k = \left\{ y_{s_{k,0}},\, y_{s_{k,1}},\, \dots,\, y_{s_{k,m_k}} \right\}$.
			\For{$j \in \left\{0, 1, \dots, m_k - 1 \right\}$}
				\State Set \texttt{initial\_state} $\leftarrow y_{s_{k,j}}$, \texttt{final\_state} $\leftarrow y_{s_{k,j+1}}$, and \texttt{simulation\_time} $\leftarrow s_{k,j+1} - s_{k,j}$.
				\While{\texttt{final\_state\_simulation} $\neq$ \texttt{final\_state}} \label{algorithm.markov.bridges.rejection.step}
					\State Simulate a path of a homogeneous MJP over \texttt{simulation\_time}, starting in \texttt{initial\_state},
					\Statex \hspace{\algorithmicindent}\hspace{\algorithmicindent}\hspace{\algorithmicindent}using $\bm{\Lambda}$.
					\State Record the simulated transition times $\tilde{s}_{k,j}$ and corresponding states $y_{\tilde{s}_{k,j}}$.
					\State Set \texttt{final\_state\_simulation} to the last state visited in the simulation.
				\EndWhile
			\EndFor
			\State \label{algorithm.markov.bridges.transform.to.inhomogeneous}Transform all simulated times to the inhomogeneous scale: $\tilde{t}_{k,j} = g_{\beta}(\tilde{s}_{k,j})$ for $j = 0, \dots, n_k$.
		\EndFor
	\end{algorithmic}
\end{algorithm*}

\begin{remark}
	The rejection step of Algorithm~\ref{algorithm.markov.bridges} (line~\ref{algorithm.markov.bridges.rejection.step}) ensures that only simulated trajectories which end in the correct final state are accepted. This guarantees that each segment is a valid sample from the corresponding Markov bridge distribution.
\end{remark}

\subsubsection{EM-like Estimation under Discrete Observations}\label{subsect:em-like.estimation}
When either $\bm{\Lambda}$ or $\beta$ is fixed, estimation of the remaining parameter based on the discretely observed trajectories $\bm X^d$ can be framed as a missing data problem. In the transformed timeline, the complete data $\bm{Y}^c$ consist of the continuous-time paths of the underlying homogeneous MJP $\bm{Y}$, which are only partially observed at the mapped discrete time points. The missing data correspond to the exact times of all state transitions and any unobserved states visited between these discrete observation points; for right-censored paths, they also include the unobserved continuation of the trajectory beyond the censoring time.

To handle this, we employ a Stochastic Expectation-Maximization (SEM) algorithm \citep{ESP}, which alternates between simulating the missing data and conditionally updating the parameters:
\begin{itemize}
    \item \textbf{SE-step:} Using the current estimate $\hat{\beta}$, we map the observation times to the homogeneous scale via $s = g^{-1}_{\hat{\beta}}(t)$ and simulate the full continuous-time trajectories using Markov bridges conditioned on the discrete observations, and generate the unobserved tails until absorption. This provides a complete-data realization upon which we can evaluate the log-likelihood.
    \item \textbf{M-step:} Using the simulated complete trajectories, we alternate between estimating $\bm{\Lambda}$ and estimating $\beta$.
    \begin{enumerate}
        \item \textbf{Estimating $\bm{\Lambda}$ given $\beta$:} Given a fixed $\beta$, we use the simulated complete trajectories mapped to the homogeneous time scale and maximize the homogeneous complete-data log-likelihood defined in Equation \eqref{loglike}. This yields the standard M-step updates for the sub-intensity matrix as detailed in Proposition \ref{prop:MLEs}.
        \item \textbf{Estimating $\beta$ given $\bm{\Lambda}$:} Given a fixed $\bm{\Lambda}$, we map the simulated completed trajectories back to the original inhomogeneous time scale using $t = g_{\hat{\beta}}(s)$, where $\hat{\beta}$ is the current estimate. We then estimate $\beta$ by maximizing the inhomogeneous complete-data log-likelihood defined in Equation \eqref{eq.LogLikelihood.Complete.Inhomogeneous}. Because a closed-form solution is generally unavailable, we rely on numerical optimization techniques, such as gradient ascent. The resulting value replaces $\hat{\beta}$ and is used for the time transformations in the next iteration.
    \end{enumerate}
\end{itemize}

Note that the current estimates determine not only the objective being maximized, but also the time transformation used to move between the two timelines. This alternating mechanism serves as the core subroutine embedded within the full joint estimation framework detailed in Algorithm~\ref{main.algorithm}.

\subsubsection{Identifiability} \label{subsect:identifiability}
Before introducing our full estimation method, we discuss an important issue in both homogeneous and inhomogeneous phase-type models. This is parameter identifiability.

In the subclass $ \boldsymbol{\Lambda}(t)=h_\beta(t)\boldsymbol{\Lambda}$,
one must in particular rule out confounding between the baseline matrix $ \boldsymbol{\Lambda}$ and the time-scaling function $h_\beta(t)$. Indeed, if two different values of $\beta$ produce scaling functions that differ only by a multiplicative constant, then this constant can be absorbed into $ \boldsymbol{\Lambda}$, yielding the same time-dependent sub-intensity matrix.

The next proposition characterizes when this type of confounding cannot occur, that is, when the parameters are identifiable.

\begin{proposition}
Consider the family of time-dependent sub-intensity matrices of the form  $ \boldsymbol{\Lambda}(t)=h_\beta(t)\boldsymbol{\Lambda}$. Then the parametrization $ \left( \boldsymbol{\Lambda}, \beta \right)$ is identifiable at the sub-intensity matrix level if and only if the following condition holds: whenever $ \beta_1$ and  $\beta_2$ satisfy
\begin{equation*}
    h_{\beta_1}(t)=c\,h_{\beta_2}(t)\qquad \text{for all } t \geq 0
\end{equation*}
for some constant $c>0$, it follows that $\beta_1=\beta_2$.
\end{proposition}

\begin{proof}
    $\Rightarrow]$ Suppose first that there exist $\beta_1\neq \beta_2 $ and a constant \(c>0\) such that
    \begin{equation*}
        h_{\beta_1}(t)=c\,h_{\beta_2}(t)
        \qquad\text{for all } t.
    \end{equation*}
    Then, for any sub-intensity matrix $\boldsymbol{\Lambda}$,
    \begin{equation*}
        h_{\beta_1}(t)\boldsymbol{\Lambda}
        =
        h_{\beta_2}(t)(c\boldsymbol{\Lambda})
        \qquad\text{for all } t.
    \end{equation*}
    Hence the two parameter pairs $\left(\boldsymbol{\Lambda},\beta_1\right)$ and $\left(c\boldsymbol{\Lambda},\beta_2\right)$ generate the same time-dependent sub-intensity matrix, so the parametrization is not identifiable.
    
    $\Leftarrow] $ Conversely, assume that the stated condition holds, and suppose that two parameter pairs $\left(\boldsymbol{\Lambda}_1,\beta_1\right)$ and $\left(\boldsymbol{\Lambda}_2,\beta_2\right)$ satisfy
    \begin{equation*}
        h_{\beta_1}(t) \boldsymbol{\Lambda}_1 = h_{\beta_2}(t) \boldsymbol{\Lambda}_2
        \qquad\text{for all } t.
    \end{equation*}
    Assuming the matrices are non-zero, we can fix at least one entry  $(i,j)$ such that $(\boldsymbol{\Lambda}_1)_{ij}\neq 0 $ and $(\boldsymbol{\Lambda}_2)_{ij}\neq 0 $. Then
    \begin{equation*}
        \frac{h_{\beta_1}(t)}{h_{\beta_2}(t)}
        =
        \frac{(\boldsymbol{\Lambda}_2)_{ij}}{(\boldsymbol{\Lambda}_1)_{ij}},
    \end{equation*}
    which is independent of $t$. Therefore, the ratio $ h_{\beta_1}(t)/h_{\beta_2}(t) $ is constant in $t$. By assumption, this implies $ \beta_1=\beta_2 $. Substituting back, we obtain
    \begin{equation*}
        h_{\beta_1}(t)\boldsymbol{\Lambda}_1 = h_{\beta_1}(t)\boldsymbol{\Lambda}_2,
    \end{equation*}
    and hence $\boldsymbol{\Lambda}_1 = \boldsymbol{\Lambda}_2$. Therefore, whenever two parameter pairs produce the same time-dependent sub-intensity matrix, the pairs must coincide. This is exactly the identifiability of the parametrization $ \left(\boldsymbol{\Lambda}, \beta \right) $.
\end{proof}
\subsection{Full Estimation Procedure} \label{subsection.full.estimation.procedure}
We now have all the necessary tools to describe our complete estimation algorithm. In the remainder of this section, we outline the initialization, iterative procedure, and termination criterion in detail.

\subsubsection{Initialization} \label{subsect:initialization}
We begin by specifying an initial guess for the parameter $\beta$, denoted by $\hat{\beta}_0$. While the exact value of $\hat{\beta}_0$ is not critical for the correctness of the algorithm, choosing a value close to the true parameter can significantly improve convergence speed.

As previously mentioned, the algorithm relies on a gradient ascent procedure. As such, it requires both a step size $\eta > 0$ and a convergence threshold $e_\ell > 0$ for the improvement in the log-likelihood. These two hyperparameters should be selected with care, as they directly influence both the convergence behavior and computational efficiency of the algorithm.

As part of the initialization, we also construct an initial estimate of the sub-intensity matrix, denoted $\boldsymbol{\hat{\Lambda}}_0$. To do this, we temporarily treat the $K$ inhomogeneous and discretely observed trajectories as if they were fully observed (i.e., continuous in time) and governed by a homogeneous process. Under these simplifying assumptions, we apply Proposition~\ref{prop:MLEs} (without performing a time transformation) and obtain $\boldsymbol{\hat{\Lambda}}_0$ using equation~\eqref{mles}. We relabel this initial estimate $\boldsymbol{\hat{\Lambda}}_0$ as $\boldsymbol{\hat{\Lambda}}$, which will serve as the starting point for the iterative procedure.

We complete the initialization step by refining our initial guess $\hat{\beta}_0$. Because the gradient ascent update requires continuous-time data, we first simulate an initial set of continuous trajectories for all $K$ paths using Markov bridges conditioned on the discrete observations and $\boldsymbol{\hat{\Lambda}}_0$. Consistent with the construction of $\boldsymbol{\hat{\Lambda}}_0$, this initial simulation is carried out directly on the original time scale, treating the observations as if they came from a homogeneous process. As no time transformation is applied, no value of $\beta$ is required at this stage. In practice this amounts to applying Algorithm~\ref{algorithm.markov.bridges} with Steps~\ref{algorithm.markov.bridges.transform.to.homogeneous} and~\ref{algorithm.markov.bridges.transform.to.inhomogeneous} omitted. 

At this stage, because our initial parameter estimates may not yet be reliable, we explicitly \textit{do not} simulate the unobserved tails (i.e., the latent absorption times) for paths that are right-censored. Instead, we only reconstruct the transitions between the observed discrete points, treating the unabsorbed paths as right-censored exactly at their final observation time. Because no transformation was applied, the simulated transition times are expressed in the original inhomogeneous timeline and can therefore be used directly in the gradient ascent update defined in Equation~\eqref{eq:GD.update} using $\boldsymbol{\hat{\Lambda}}_0$, $\eta$, and $e_\ell$. This update is repeated until the log-likelihood convergence criterion specified in Equation~\eqref{eq:GD.convergence.criteria} is met, yielding a refined initial estimate $\hat{\beta}$ to begin the main SEM iterations.

\subsubsection{Iterative Procedure} \label{subsect:iterative.procedure}
The following steps form an SEM-like iterative procedure:  each iteration begins by simulating the latent continuous-time paths (SE-step), followed by two conditional maximization steps (M-steps): updating $\bm{\hat{\Lambda}}$ conditional on $\hat{\beta}$, and refining $\hat{\beta}$ conditional on $\hat{\bm{\Lambda}}$. These steps are repeated until the termination criterion described in Section~\ref{subsect:termination.criteria} is satisfied.

We begin by transforming all originally observed inhomogeneous time points $\left\{t_{k,0},\, t_{k,1},\, \dots,\, t_{k,m_k} \right\}_{k=1}^K$ into their homogeneous counterparts. This is done using the time transformation given in equation~\eqref{eq:Connection.Inhomogeneous.Homogeneous} with the most recent $\hat{\beta}$, yielding the transformed time points:
\begin{multline*}
	\left\{ s_{k,0} = g^{-1}_{\hat{\beta}}\left(t_{k,0}\right),\, s_{k,1} = g^{-1}_{\hat{\beta}}\left(t_{k,1}\right),\, \dots,\,\right. \\
	\left. s_{k,m_k} = g^{-1}_{\hat{\beta}}\left(t_{k,m_k}\right) \right\}_{k=1}^K.
\end{multline*}

Using these transformed data, we simulate continuous-time trajectories in the homogeneous timeline via Markov bridges, as described in Section~\ref{subsect:handling.discrete.times} and, more specifically, in Algorithm~\ref{algorithm.markov.bridges}.

In applying Algorithm~\ref{algorithm.markov.bridges}, we skip Step~\ref{algorithm.markov.bridges.transform.to.homogeneous} since the time transformation has already been performed, and we also skip Step~\ref{algorithm.markov.bridges.transform.to.inhomogeneous}, as we will revert to the original time scale later. The simulated continuous-time data in the homogeneous timeline is denoted by:
\begin{equation*}
	\left\{ y_{\tilde{s}_{k,0}},\, y_{\tilde{s}_{k,1}},\, \dots,\, y_{\tilde{s}_{k,n_k}} \right\}_{k=1}^K,
\end{equation*}
where $\tilde{s}_{k,j}$ denotes the latent state transition times.%
\footnote{We continue to use the convention from Remark~\ref{rmk.discrete.vs.continuous.times} to distinguish between observed discrete times and latent continuous transitions.}

Special care must be taken for paths that are not absorbed by time $S= g_{\hat{\beta}}^{-1} \left(T\right)$. Suppose path $k$ is such a case. Then the final simulated time $\tilde{s}_{k,n_k}$ satisfies $\tilde{s}_{k,n_k} < s_{k,m_k}$, implying that the last observed data point at $s_{k,m_k}$ is not accounted for in the reconstructed path. To retain this information, we add the observation at the final time point to the simulated trajectory:
\begin{equation*}
	\left\{ y_{\tilde{s}_{k,0}},\, y_{\tilde{s}_{k,1}},\, \dots,\, y_{\tilde{s}_{k,n_k}},\, y_{s_{k,m_k}} \right\}.
\end{equation*}

Next, for each non-absorbed path, we simulate an absorption time. Specifically, we generate a homogeneous MJP trajectory using the current estimate $\boldsymbol{\hat{\Lambda}}$, starting from the final observed state $y_{s_{k,m_k}}$, and continue the simulation until the process reaches the absorbing state. This step ensures that each path, whether originally absorbed or censored, is completed as a full homogeneous trajectory terminating in absorption. For paths that are censored in the observed data, the simulated terminal absorption is treated as part of the augmented complete trajectory generated in the SE-step.

For the first part of the M-step, we update the estimate $\boldsymbol{\hat{\Lambda}}$. To do so, we apply equation~\eqref{mles} from Proposition~\ref{prop:MLEs} in the homogeneous timeline, using all available simulated continuous-time data. This includes both the reconstructed trajectories between observation points (obtained via Markov bridges) and the extended simulations until absorption for originally non-absorbed paths.

For the second part of the M-step, we refine the estimate of $\hat{\beta}$. We collect the entire simulated continuous-time trajectories (including all intermediate transition times and holding intervals) from the homogeneous timeline and map them back to the inhomogeneous time scale using the transformation $t = g_{\hat{\beta}}(s)$, based on the current estimate of $\hat{\beta}$. Using these fully reconstructed inhomogeneous paths, we apply the gradient ascent update defined in equation~\eqref{eq:GD.update} to maximize the inhomogeneous complete-data log-likelihood. This update utilizes the current estimate of $\boldsymbol{\hat{\Lambda}}$, the step size $\eta$, and is repeated until the change in the proportional log-likelihood falls below the convergence threshold $e_\ell$, as specified in equation~\eqref{eq:GD.convergence.criteria}.

\subsubsection{Termination Criterion} \label{subsect:termination.criteria}
As noted at the end of Section~\ref{subsect:iterative.procedure}, each outer iteration of the algorithm may involve multiple updates of the parameter estimate $\hat{\beta}$ before the convergence condition in equation~\eqref{eq:GD.convergence.criteria} is met. Accordingly, we define the termination criterion as follows: the iterative procedure terminates the first time an outer iteration requires only a single gradient-ascent update of $\hat{\beta}$ to satisfy the convergence criterion.

This stopping rule captures the idea that the estimates have sufficiently stabilized. {Indeed, if the inner gradient-ascent procedure for refining $\hat{\beta}$ satisfies the convergence criterion after a single update, then the complete-data log-likelihood is already nearly stable with respect to $\beta$ for the current simulated complete trajectory. This indicates that further outer iterations are unlikely to produce significant changes.

To bring all components together, the full estimation procedure begins with parameter initialization and proceeds via an iterative SEM-like structure that alternates between simulating continuous paths, refining estimates through likelihood maximization and gradient ascent, and transitioning between the inhomogeneous and homogeneous timelines. The complete procedure is summarized below in Algorithm~\ref{main.algorithm}.

\begin{algorithm*}[htbp]
	\caption{Summary of the Full Estimation Procedure}
	\label{main.algorithm}
	\textbf{Input:} Discrete-time observations $\bm{X}^d$, initial guess $\hat{\beta}_0$, step size $\eta > 0$, convergence threshold $e_\ell > 0$\\
	\textbf{Output:} Estimated parameters $\hat{\bm{\pi}},\, \hat{\bm{\Lambda}},\, \hat{\beta}$
	
	\begin{algorithmic}[1]
		\State \textbf{Initialization:}
		\State Estimate initial distribution $\boldsymbol{\hat{\pi}}$ using empirical frequencies (equation~\eqref{estimation.pi}).
		\State Estimate $\boldsymbol{\hat{\Lambda}}_0$ by treating $\boldsymbol{X}^d$ as continuous (no time transformation) and applying equation~\eqref{mles}.
		\State Set $\bm{\hat{\Lambda}} \leftarrow \bm{\hat{\Lambda}}_0$.
		\State Identify all paths absorbed before time $T$.
		\For{each path $k = 1, \dots, K$}
		\State Simulate latent continuous-time transitions between observed points using Markov bridges
		\Statex \hspace{\algorithmicindent}(Algorithm~\ref{algorithm.markov.bridges}, omitting Steps~\ref{algorithm.markov.bridges.transform.to.homogeneous} and~\ref{algorithm.markov.bridges.transform.to.inhomogeneous}) and $\boldsymbol{\hat{\Lambda}}$.
		\If{path $k$ is not absorbed}
		    \State Add the last observed state $y_{t_{k,m_k}}$ to the trajectory (do \textit{not} simulate absorption time yet).
		\EndIf
	   \EndFor
		\State Set $\hat{\beta} \leftarrow \hat{\beta}_0$
		\Repeat
			\State Refine $\hat{\beta}$ via gradient ascent (equation~\eqref{eq:GD.update}) using $\boldsymbol{\hat{\Lambda}}$, $\eta$, and $e_\ell$.
		\Until{convergence criterion~\eqref{eq:GD.convergence.criteria} is satisfied}.
		\State \textbf{Iterative Estimation:}
		\Repeat
			\State Transform all observed time points to the homogeneous scale using $s = g^{-1}_{\hat{\beta}}(t)$.
			\For{each path $k = 1, \dots, K$}
				\State Simulate latent continuous-time transitions between $s_{k,0}, \dots, s_{k,m_k}$ using Markov bridges
				\Statex \hspace{\algorithmicindent}\hspace{\algorithmicindent} (Algorithm~\ref{algorithm.markov.bridges})
				\If{path $k$ is not absorbed by time $S = g^{-1}_{\hat{\beta}}(T)$}
					\State Add the last observed state $y_{s_{k,m_k}}$ to the simulated trajectory.
					\State Simulate absorption time (in homogeneous time) starting from $y_{s_{k,m_k}}$ using $\boldsymbol{\hat{\Lambda}}$.
				\EndIf
			\EndFor
			\State Update $\boldsymbol{\hat{\Lambda}}$ via MLE using equation~\eqref{mles} and all simulated homogeneous trajectories.
			\State Transform all simulated continuous-time trajectories (transitions and holding intervals) 
		      \Statex \hspace{\algorithmicindent} from the homogeneous to the inhomogeneous scale using $t = g_{\hat{\beta}}(s)$.    
			\Repeat
				\State Update $\hat{\beta}$ via gradient ascent (equation~\eqref{eq:GD.update}) using $\boldsymbol{\hat{\Lambda}}$, $\eta$, and $e_\ell$.
			\Until{convergence criterion~\eqref{eq:GD.convergence.criteria} is satisfied}
		\Until{an iteration in which criterion~\eqref{eq:GD.convergence.criteria} is satisfied after a single update of $\hat\beta$.}
		\State \textbf{Return} $\hat{\bm{\pi}},\, \bm{\hat{\Lambda}},\, \hat{\beta}$
	\end{algorithmic}
\end{algorithm*}
\begin{remark}
	Given the current estimate of $\hat{\beta}$, Steps 18-26 in Algorithm~\ref{main.algorithm} implement a SEM update for $\bm \Lambda$; under mild regularity conditions, SEM iterations are known to increase the complete-data likelihood on average and to converge in distribution toward a neighborhood of a maximum of the likelihood \citep{Kri}.
	
	Similarly, Steps 28-30 apply gradient ascent to the complete-data log-likelihood in $\beta$. With an appropriate choice of step size $\eta$ and convergence threshold $e_\ell$, this procedure typically converges to a local maximizer of the complete-data objective \citep{bottou2018optimization}.
	
	Overall, for a fixed simulated complete-data realization, the algorithm alternates between updating $\bm \Lambda$ and $\beta$, and both updates are exact maximizations of the complete-data log-likelihood for that realization: the update for $\bm \Lambda$ is the closed-form complete-data MLE of Proposition~\ref{prop:MLEs}, and the update for $\beta$ maximizes equation \eqref{eq:log.lik.proportional.beta} numerically. Since the missing trajectories are resampled at each SEM iteration, the likelihood need not be monotone across outer iterations and stochastic fluctuations may occur. In our experiments, this SEM-GA scheme consistently stabilized after a moderate number of iterations, yielding parameter estimates that provided excellent fits to both simulated and real datasets.

    It is worth emphasizing what the algorithm returns. The output is not the maximizer of the observed-data likelihood, but a complete-data maximum-likelihood estimator evaluated at a simulated realization of the latent trajectories drawn under the current parameter values. Every iteration thus produces the MLEs associated with a simulated complete dataset, and the reported estimates are obtained entirely from such complete-data MLE updates, without any direct optimization of the observed-data likelihood. Under the usual SEM conditions, the sequence of these estimators forms an ergodic Markov chain whose stationary distribution concentrates near the observed-data maximizer \citep{nielsen2000stochastic}, so that the reported estimates retain the interpretation of maximum-likelihood estimates up to Monte Carlo variability.
\end{remark}

\section{Simulation Study} \label{Sect.Simulation.Study}
In this section, we conduct a simulation study using two scaling functions to evaluate the accuracy of the calibration method proposed in this work.
\subsection{Gompertz Distribution}\label{Subsect.Gompertz.Distr}
In this case, the scaling function is $h_{\beta}(t) = e^{\beta t}$, with \( \beta > 0 \). The function $ g_\beta^{-1} $ associated with this scaling function is given by
$$
g_\beta^{-1}(t) = \int_0^t e^{\beta s} \, ds = \frac{1}{\beta} \left( e^{\beta t} - 1 \right).
$$
Inverting this expression, we obtain
$$
g_\beta(s) = \frac{1}{\beta} \log(\beta s + 1).
$$
Therefore, the absorption time \( \tau \) of the inhomogeneous process satisfies
\[
\tau = \frac{1}{\beta} \log(\beta \rho + 1),
\]
where \( \rho \sim \mathrm{PH}(\bm{\pi}, \bm{\Lambda}) \). The resulting distribution of $\tau$ is known as \emph{matrix-Gompertz distribution}. The Gompertz distribution is commonly used to model human lifetimes \citep{gompertz1825}. In \cite{Al-bla-ys-20}, the authors use this distribution to model the lifetime of the Danish population. 

The distribution function of $\tau$ is given by
$$F_\tau(x; \bm \pi, \bm \Lambda,\beta) = 1 - \bm \pi e^{\bm \Lambda(e^{\beta x}-1)/\beta}\mathbbm{1}_n ,$$
and the corresponding density function is 
$$ f_\tau(x;\bm \pi, \bm \Lambda,\beta) = \bm {\pi} e^{\bm \Lambda(e^{\beta x}-1)/\beta}\bm\lambda e^{\beta x}.$$

Under the SEM-GA framework introduced in Section \ref{sect.Statistical.Inference}, given the current estimate $\hat{\bm{\Lambda}}$, the time-scaling parameter $\beta$ is updated by maximizing the complete-data log-likelihood based on the simulated continuous paths $\bm{X}^c$.

Substituting the Gompertz scaling function $h_\beta(t) = e^{\beta t}$ into the general complete-data log-likelihood on the inhomogeneous scale, and isolating the terms involving $\beta$ (see equation~\eqref{eq:log.lik.proportional.beta}), the objective function simplifies to:
\begin{multline}\label{eq:Gompertz_Likelihood_Beta}
    \ell^c(\beta) \propto \beta \sum_{k=1}^K \sum_{j=1}^{n_k} \tilde{t}_{k,j} \\
     + \sum_{x=1}^n \hat{\lambda}_{xx} \int_0^T \mathbbm{1}_{\{X_t = x\}} e^{\beta t} dt,
\end{multline}
where $\tilde{t}_{k,j}$ are the exact transition times in the reconstructed paths, and the integral represents the $\beta$-scaled occupation time in each transient state $x$.

To perform the gradient ascent update step, we need the derivative of $\ell^c(\beta)$ with respect to $\beta$:
\begin{multline} \label{eq:Gompertz_Gradient_Beta}
    \frac{d}{d\beta}\ell^c(\beta) = \sum_{k=1}^K \sum_{j=1}^{n_k} \tilde{t}_{k,j} \\
    + \sum_{x=1}^n \hat{\lambda}_{xx} \int_0^T \mathbbm{1}_{\{X_t = x\}} t e^{\beta t} dt.
\end{multline}
During the iterative procedure, $\hat{\beta}$ is refined by taking steps in the direction of this gradient until the convergence criterion is achieved.

To assess the performance of our estimation procedure, we generate $K=1000$ sample paths from an IMJP $\bm X$, defined over the time interval $[0,60]$, with $n=3$ transient states. The parameters used to generate the simulated data correspond to the estimators reported in \cite{Al-bla-ys-20}: $\bm\pi=(0.0451, 0.1303, 0.8246)$, $\beta=0.1019$, and 
\[
\bm \Lambda=
\begin{pmatrix}
	-0.1357 & 0.1214 & 0.0000 \\
	0.0130 & -0.0421 & 0.0288 \\
	0.1415 & 0.0184 & -0.1620
\end{pmatrix}.
\]

Although each path is simulated in continuous time, so that the exact transition and absorption times are known to us, we retain only the states visited at an equally spaced grid of observation times, with $t_{k,i} - t_{k,i-1} = 1$, and treat all transition and absorption times as unobserved. This discretization reproduces the panel-data observation scheme of Section~\ref{subsect:observation.scheme} and produces the discrete-time observations $\bm X^d$, which we treat as the observed data and which are the only input to the estimation procedure. For the truncated horizons reported below, we additionally retain only those observations with $t_{k,i} \leq T$. We apply the full estimation procedure described in Algorithm~\ref{main.algorithm}, which jointly estimates the parameters $(\bm{\hat{\pi}}, \bm{\hat{\Lambda}}, \hat{\beta})$ from these discretely observed sample paths.

The SEM algorithm is initialized, using the method described in Section \ref{subsect:initialization}, with
\begin{itemize}
	\item The initial value $\hat{\beta}_0 = 1$,
	\item An initial sub-intensity matrix $\bm{\hat{\Lambda}}_0$, obtained by applying the maximum likelihood estimator for homogeneous MJPs described in Proposition~\ref{prop:MLEs}, treating the discrete-time observations $\bm{X}^d$ as if they were continuously observed paths from a homogeneous process (i.e., no time transformation was applied).
\end{itemize}

At each iteration, as detailed in Section \ref{subsect:iterative.procedure}, continuous-time sample paths are reconstructed using Markov bridges, followed by maximum likelihood updates for $\bm{\hat{\Lambda}}$ and a gradient-based refinement of $\hat{\beta}$, specifically using the update rule given in equation~\eqref{eq:GD.update} with learning rate $\eta = 10^{-8}$ and convergence threshold $e_\ell = 0.01$.

In Table~\ref{tab:estimators-Gompertz}, we present the estimated values of the parameters for different time horizons $T$. Recall that, as defined in Section~\ref{sect.Statistical.Inference}, $T$ represents the total observation time.

\begin{table*}[htbp]
	\centering
	\caption{Estimated Parameters for Different Values of $T$ (Gompertz Model)}
	\label{tab:estimators-Gompertz}
	\small
	\begin{tabular}{ccccccccccc}
		\toprule
		\( T \) & \( \hat{\beta} \) & \( \hat{\lambda}_{11} \) & \( \hat{\lambda}_{12} \) & \( \hat{\lambda}_{13} \) & \( \hat{\lambda}_{21} \) & \( \hat{\lambda}_{22} \) & \( \hat{\lambda}_{23} \) & \( \hat{\lambda}_{31} \) & \( \hat{\lambda}_{32} \) & \( \hat{\lambda}_{33} \) \\
		\midrule
		60 & 0.0966 & -0.1430 & 0.1262 & 0.0002 & 0.0124 & -0.0442 & 0.0310 & 0.1572 & 0.0154 & -0.1742 \\
		47 & 0.0955 & -0.1435 & 0.1268 & 0.0001 & 0.0112 & -0.0438 & 0.0316 & 0.1558 & 0.0156 & -0.1735 \\
		39 & 0.1145 & -0.1137 & 0.1018 & 0.0000 & 0.0090 & -0.0341 & 0.0250 & 0.1285 & 0.0141 & -0.1510 \\
		36 & 0.1426 & -0.0796 & 0.0709 & 0.0013 & 0.0063 & -0.0236 & 0.0173 & 0.0923 & 0.0123 & -0.1051 \\
		\midrule
		\makecell[c]{\textbf{True} \\ \textbf{Values}} & \textbf{0.1019} & \textbf{-0.1357} & \textbf{0.1214} & \textbf{0.0000} & \textbf{0.0130} & \textbf{-0.0421} & \textbf{0.0288} & \textbf{0.1415} & \textbf{0.0184} & \textbf{-0.1620} \\
		\bottomrule
	\end{tabular}
\end{table*}

Since our primary interest is estimating the time-inhomogeneous sub-intensity matrix, we omit the estimates of the initial distribution $ \bm{\pi} $. These estimates are not central to our algorithm and, importantly, remain constant across all values of $T$, as they depend solely on the distribution of initial states at time $0$.

Shorter observation windows lead to fewer absorption events in the dataset, limiting the available information and reducing the accuracy of the estimated parameters. This pattern is reflected in Table~\ref{tab:simulation-results-Gompertz}, which reports the number of absorbed paths (out of 1000), the iteration at which the algorithm converged, and a p-value assessing the fit for each value of $ T $. 

The p-values are computed using a two-sample Kolmogorov-Smirnov (KS) test applied to exact absorption times. Recall that the data used for estimation are obtained by simulating $K = 1000$ complete trajectories under the true parameters $\left(\pi, \bm \Lambda, \beta\right)$, discretizing them, and retaining only the observations up to the horizon $T$. The KS test, by contrast, compares two untruncated samples of $1000$ absorption times each: the first is generated independently under the true parameters $\left(\bm\pi, \bm \Lambda, \beta\right)$ and run until absorption, and the second is generated under the estimates $\left(\hat{\bm \pi}, \hat{\bm \Lambda}, \hat{\beta}\right)$ obtained by applying Algorithm~\ref{main.algorithm} to the truncated data, also run until absorption. The test therefore assesses the absorption-time distribution implied by the fitted model against the true one, rather than against the truncated sample used to fit it. A larger p-value indicates less evidence against the null hypothesis that the two samples come from the same distribution, whereas a small p-value suggests that the fitted model fails to reproduce the true absorption-time distribution.

\begin{table}[htbp]
	\centering
	\caption{Simulation Results and Goodness-of-Fit for Different Values of $T$ (Gompertz Model)}
	\label{tab:simulation-results-Gompertz}
	\begin{tabular}{cccc}
		\toprule
		$T$ & Absorbed Paths & Iteration & p-value \\
		\midrule
		60 & 1000 & 67 & 0.3700 \\
		47 & 983 & 66 & 0.4324 \\
		39 & 834 & 89 & 0.4658 \\
		36 & 747 & 207 & $8.914 \times 10^{-6}$\\
		\bottomrule
	\end{tabular}
\end{table}

When $ T = 36 $, the observed time covers only about 60\% of the full trajectory horizon, leading to a significant loss of absorption information. This loss explains the extremely low p-value in the last row of Table~\ref{tab:simulation-results-Gompertz}, indicating that the estimated distribution diverges substantially from the true one in that setting.

This behavior is further illustrated in Figure~\ref{fig:absorption-distributions}, which compares the empirical distributions of the observed absorption times (under the true parameters) with those of the simulated absorption times (under the estimated parameters). For large values of $ T $, the two distributions are in very close agreement confirming the quality of the fit. As $ T $ decreases, noticeable discrepancies emerge, consistent with the sharp drop in the p-value at $T=36$. In particular, the empirical distribution of absorption times becomes increasingly left-skewed and compressed, indicating that shorter observation windows not only reduce the number of observed absorption events, but also systematically truncate the right tail of the distribution.

\begin{figure*}[htbp]
	\centering
	\includegraphics[width=0.45\textwidth]{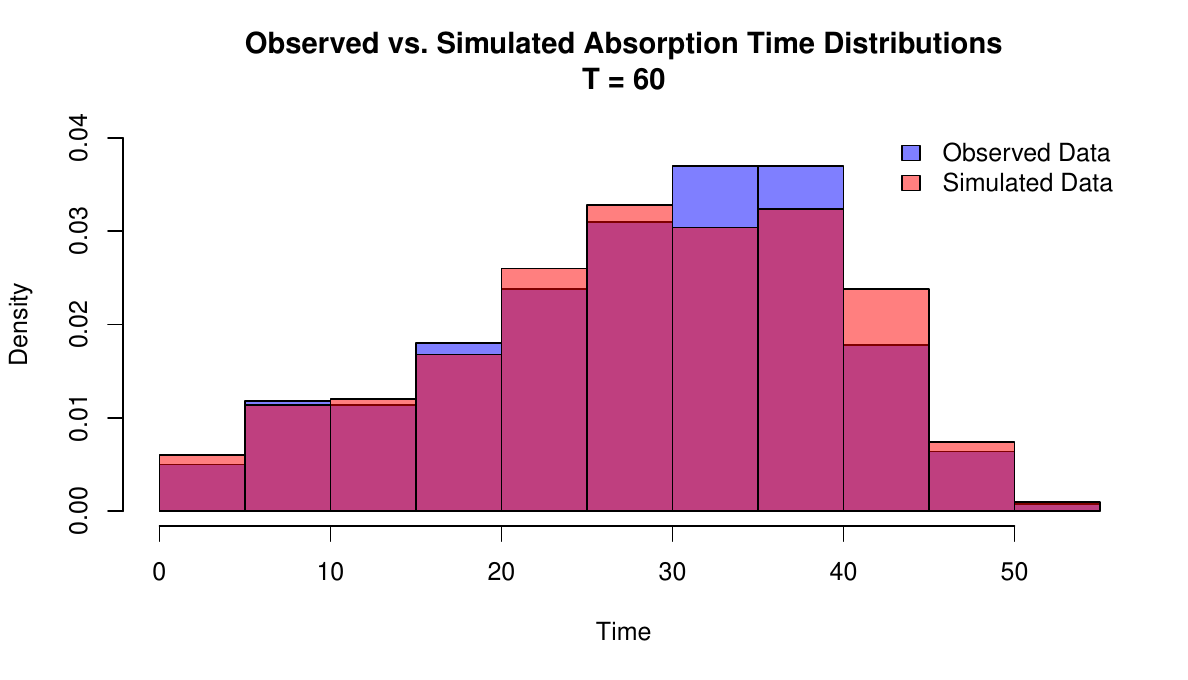}
	\includegraphics[width=0.45\textwidth]{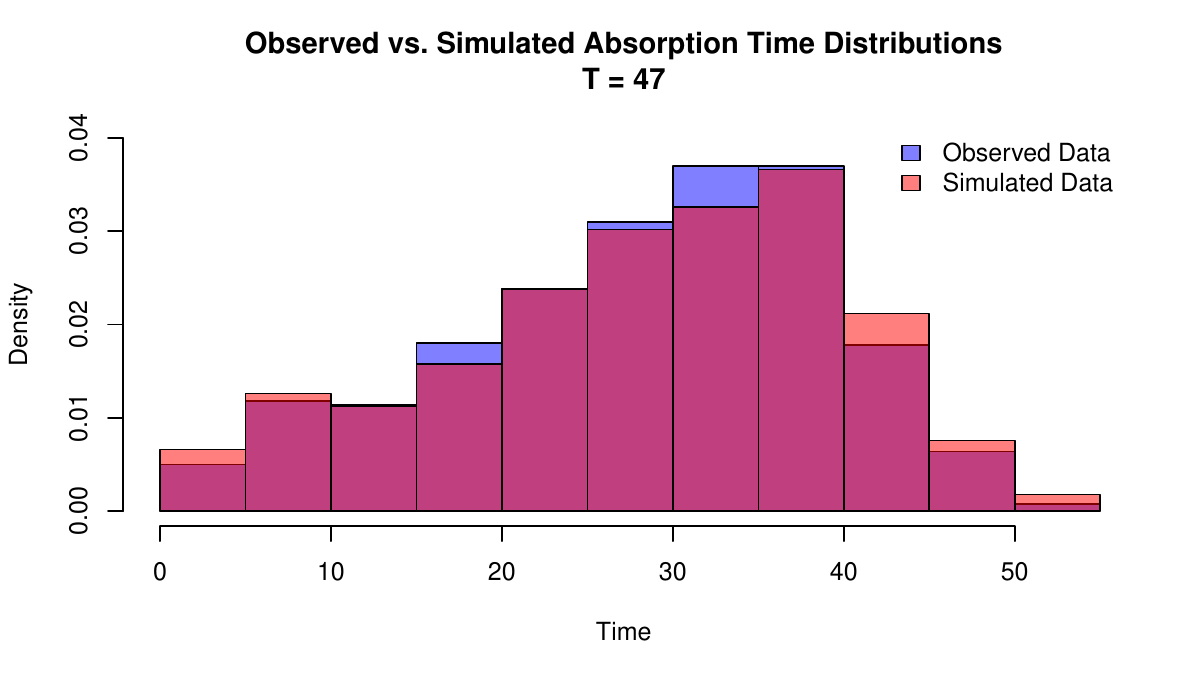}\\
	\includegraphics[width=0.45\textwidth]{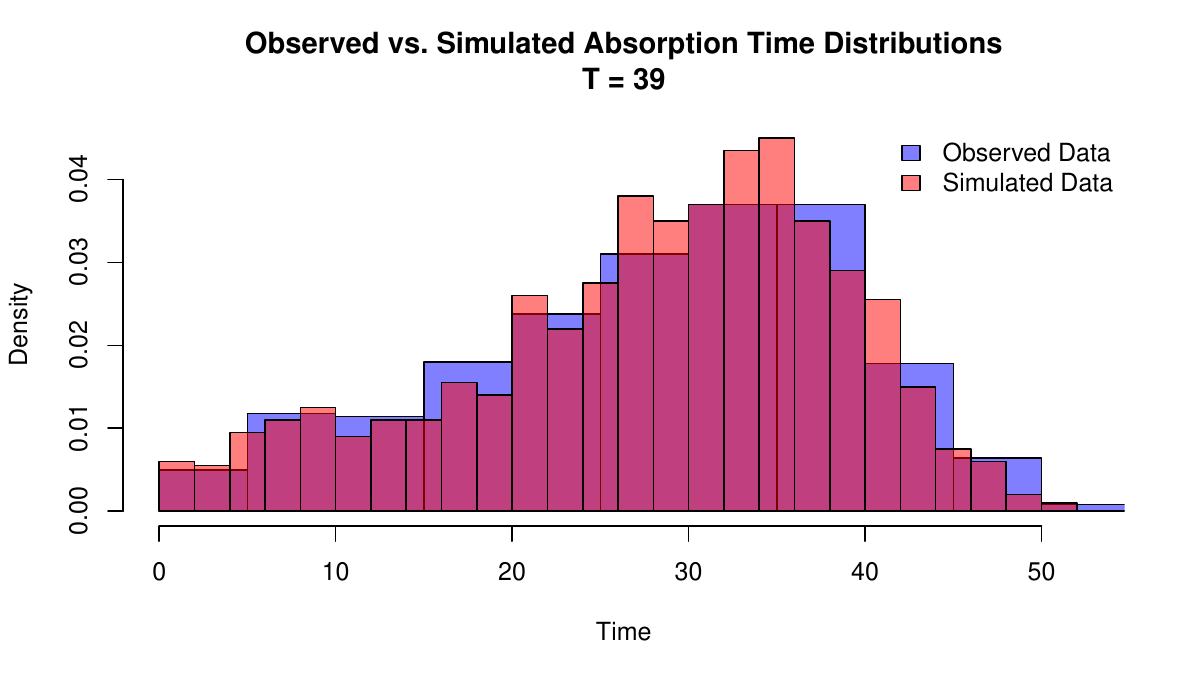}
	\includegraphics[width=0.45\textwidth]{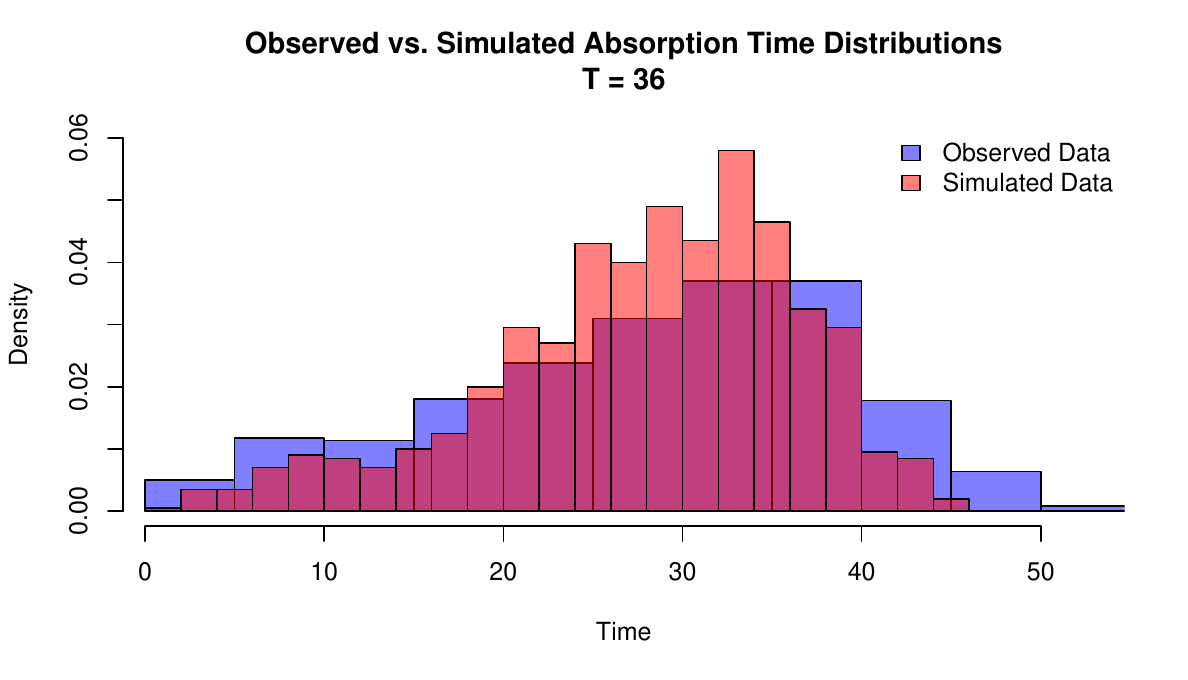}
	\caption{Comparison of absorption times generated under the true parameters with absorption times generated under the parameters estimated from data truncated at $T$. Each panel corresponds to a different value of $T$. (Gompertz Model)}
	\label{fig:absorption-distributions}
\end{figure*}

\subsection{Matrix Weibull Distribution}
In this case, the scaling function is defined as $h_\beta(t) = \beta t^{\beta - 1}$, with $\beta > 0$. The associated transformation $g_\beta^{-1}$ is given by:
\[
g_\beta^{-1}(t) = \int_0^t \beta s^{\beta - 1} ds = t^\beta,
\]
so its inverse is
\[
g_\beta(s) = s^{1/\beta}.
\]
Therefore, if $\rho \sim \mathrm{PH}(\pi, \Lambda)$, then the absorption time of the inhomogeneous process satisfies
\[
\tau = \rho^{1/\beta},
\]
and the resulting distribution of $\tau$ is referred to as matrix Weibull distribution. It is defined through a matrix transformation of the Weibull cumulative distribution function, enabling flexible modeling of lifetimes and survival data with complex dependence structures and non-monotonic hazard rates \citep{rojas}. 

The density and distribution functions of $\tau$ are given by:
\begin{align*}
	F_\tau(x; \pi, \Lambda, \beta) &= 1 - \bm \pi \, e^{\bm \Lambda x^\beta} \mathbbm{1}_n, \\
	f_\tau(x; \pi, \Lambda, \beta) &= \bm \pi \, e^{\bm \Lambda x^\beta} \bm \lambda \cdot \beta x^{\beta - 1}.
\end{align*}

Substituting the Weibull scaling function $h_\beta(t) = \beta t^{\beta-1}$ into the general complete-data log-likelihood on the inhomogeneous scale, and isolating the terms involving $\beta$ (see equation~\eqref{eq:log.lik.proportional.beta}), the objective function simplifies to:
\begin{multline} \label{eq:Weibull_Likelihood_Beta}
    \ell^c(\beta) \propto N_{\text{tot}} \log \beta + (\beta - 1) \sum_{k=1}^K \sum_{j=1}^{n_k} \log(\tilde{t}_{k,j}) \\ 
    + \sum_{x=1}^n \hat{\lambda}_{xx} \int_0^T \mathbbm{1}_{\{X_t = x\}} \beta t^{\beta-1} dt,
\end{multline}
where $N_{\text{tot}} = \sum_{k=1}^K n_k$ is the total number of observed transitions across all $K$ paths in the reconstructed data, $\tilde{t}_{k,j}$ are the exact transition times, and the integral represents the scaled occupation time in transient state $x$.

To perform the gradient ascent update step, we evaluate the derivative of $\ell^c(\beta)$ with respect to $\beta$:
\begin{multline} \label{eq:Weibull_Gradient_Beta}
    \frac{d}{d\beta}\ell^c(\beta) = \frac{N_{\text{tot}}}{\beta} + \sum_{k=1}^K \sum_{j=1}^{n_k} \log\left(\tilde{t}_{k,j}\right) \\
    + \sum_{x=1}^n \hat{\lambda}_{xx} \int_0^T \mathbbm{1}_{\{X_t = x\}} t^{\beta-1}(1 + \beta \log t) dt.
\end{multline}
During the iterative procedure, $\hat{\beta}$ is refined by taking steps in the direction of this gradient until the convergence criterion is met.

To assess the performance of the estimation procedure, we generate $K = 1000$ sample paths from an IMJP $X$ on the interval $[0,5]$ with $n = 2$ transient states. The parameters used are taken from \cite{BladtPeralta2025}:
\[
\bm{\pi} = (0.5, 0.5), \quad \beta = 3, \quad \bm{\Lambda} =
\begin{pmatrix}
	-3 & 0.1 \\
	0.01 & -0.1
\end{pmatrix}.
\]

As in Section~\ref{Subsect.Gompertz.Distr}, the paths are simulated in continuous time but only their states, at an equally spaced grid of observation times, are retained, here with $t_{k,i} - t_{k,i-1} = 0.1$ over $\left[ 0,5 \right]$ for $k = 1, \dots, K$ and $i = 1, \dots, m_k$; the exact transition and absorption times are discarded and treated as missing. The resulting discrete-time observations $\bm X^d $ are the only input to the estimation procedure. We apply the full estimation procedure described in Algorithm~\ref{main.algorithm}, which jointly estimates the parameters $(\bm{\hat{\pi}}, \bm{\hat{\Lambda}}, \hat{\beta})$.

The SEM algorithm is initialized, using the method described in Section~\ref{subsect:initialization}, with:
\begin{itemize}
	\item An initial value $\hat{\beta}_0 = 2$,
	\item An initial sub-intensity matrix $\bm{\hat{\Lambda}}_0$, obtained by applying the maximum likelihood estimator for homogeneous MJPs described in Proposition~\ref{prop:MLEs}, treating the discrete-time observations $\bm{X}^d$ as if they were continuously observed paths from a homogeneous process (i.e., no time transformation was applied).
\end{itemize}

At each iteration, as detailed in Section~\ref{subsect:iterative.procedure}, continuous-time sample paths are reconstructed using Markov bridges, followed by maximum likelihood updates for $\bm{\hat{\Lambda}}$ and a gradient-based refinement of $\hat{\beta}$, specifically using the update rule given in equation~\eqref{eq:GD.update} with learning rate $\eta = 10^{-3}$and convergence threshold $e_\ell = 10^{-3}$.

Table~\ref{tab:estimators-Weibull} presents the estimated values of the parameters at iteration 23, which corresponds to the point at which Algorithm~\ref{main.algorithm} reaches convergence. Figure~\ref{fig:evolution-Weibull} illustrates the evolution of the estimators across the 23 iterations.

The goodness-of-fit of the estimated model is assessed using the same two-sample Kolmogorov–Smirnov (KS) test described in Section \ref{Subsect.Gompertz.Distr}: we compare $1000$ absorption times generated under the true parameters $\left(\bm \pi, \bm \Lambda, \beta\right)$ with $1000$ absorption times generated under the estimates $\left(\hat{\bm \pi}, \hat{\bm \Lambda}, \hat{\beta}\right)$, all trajectories being run until absorption in both cases. The resulting p-value is 0.5004, indicating no statistically significant departure between the empirical distribution of absorption times under the estimated parameters and the distribution generated under the true parameters.
\begin{table}[htbp]
	\centering
	\caption{Estimated Parameters (Matrix Weibull Model)}
	\label{tab:estimators-Weibull}
	\begin{tabular}{ccr}
		\toprule
		\textbf{Parameter} & \textbf{True Value} & \textbf{Estimator} \\
		\midrule
		$\beta$         &   3 &   2.97204 \\
		$\lambda_{11}$  &  -3 &  -2.86971 \\
		$\lambda_{12}$  &   0.1 &  0.10194 \\
		$\lambda_{21}$  &   0.01 &   0.01238 \\
		$\lambda_{22}$  &  -0.1 &  -0.09679 \\
		\bottomrule
	\end{tabular}
\end{table}

\begin{figure*}[htbp]
	\centering
	\includegraphics[width=0.45\textwidth]{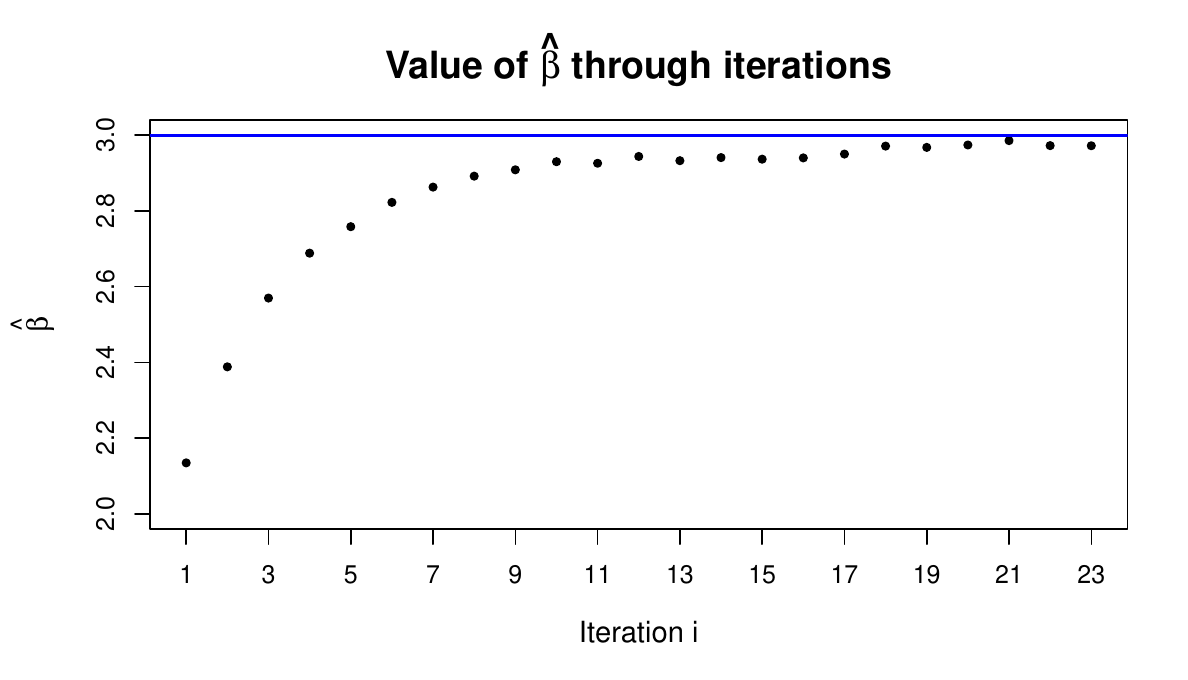}\\
	\includegraphics[width=0.45\textwidth]{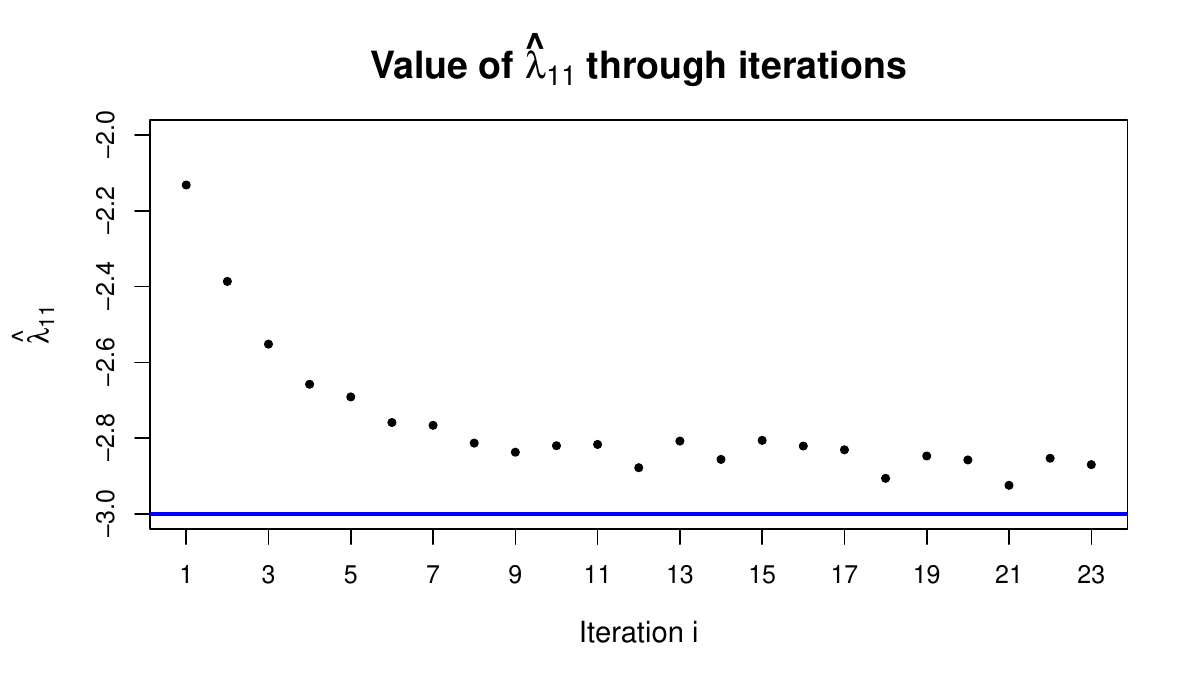}
	\includegraphics[width=0.45\textwidth]{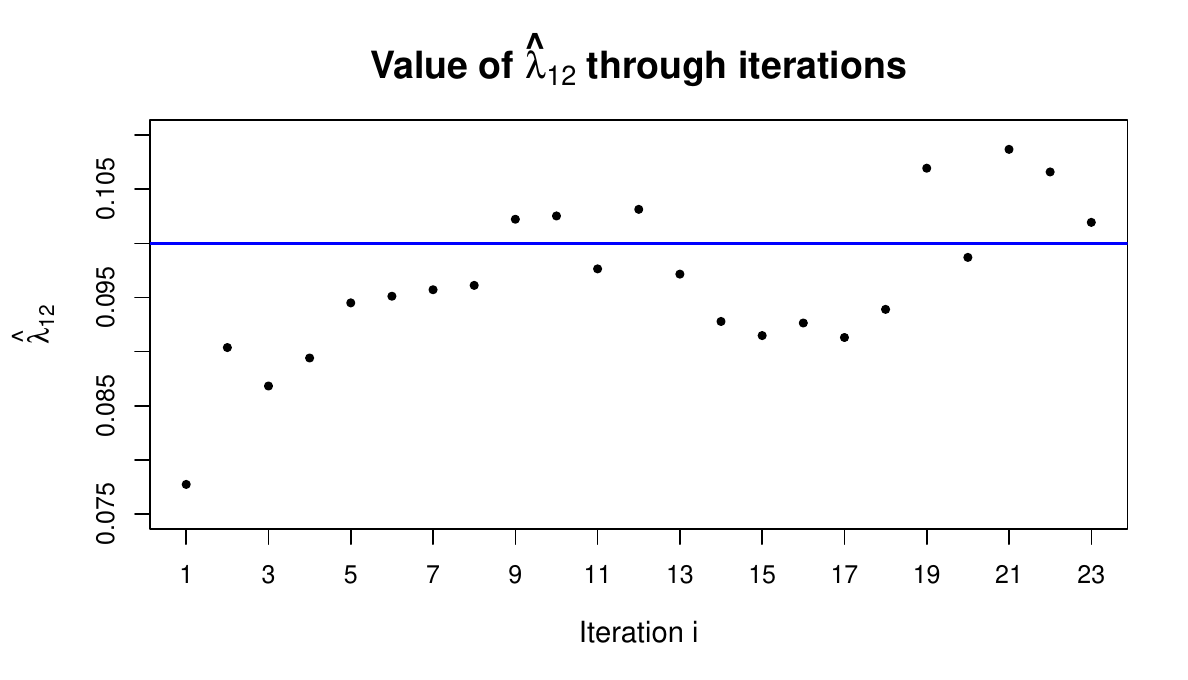}\\
	\includegraphics[width=0.45\textwidth]{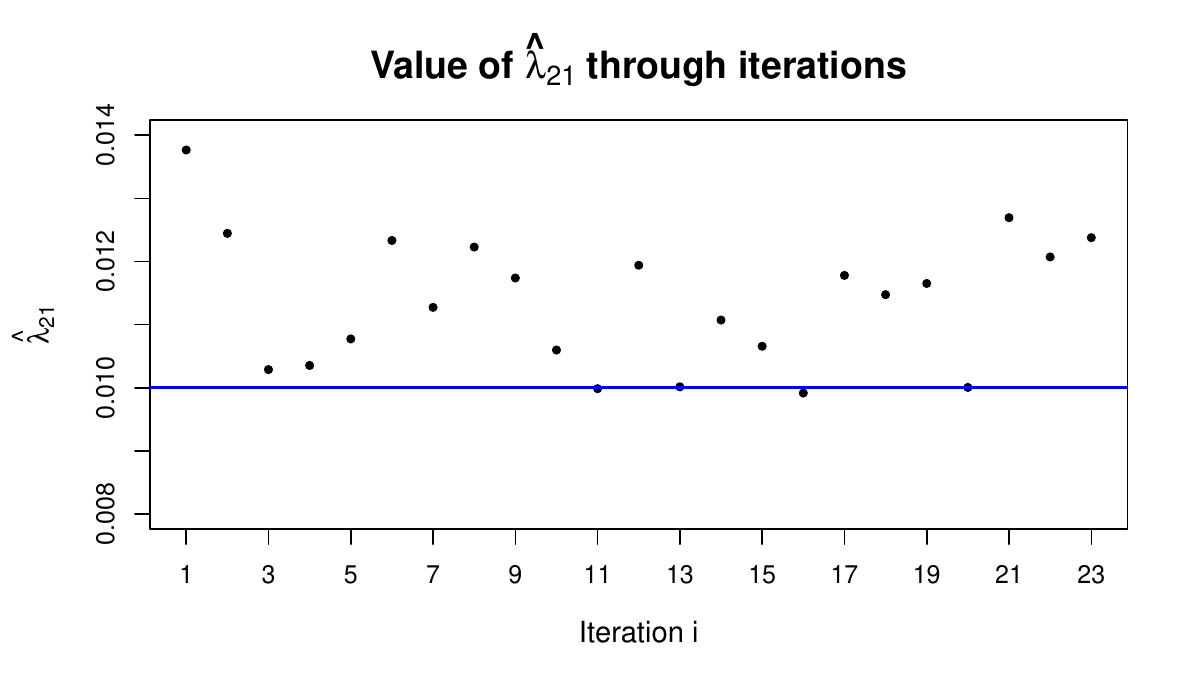}
	\includegraphics[width=0.45\textwidth]{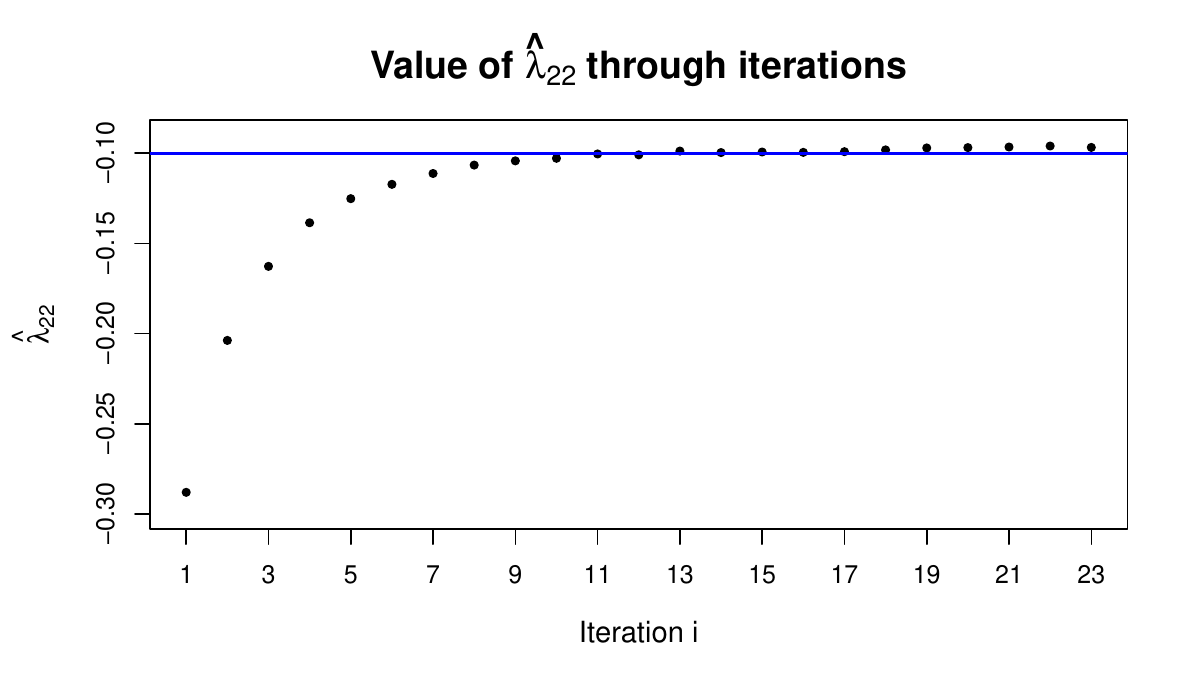}
	\caption{Evolution of the estimators across iterations, converging at iteration 23. The horizontal blue line indicates the true parameter value. Note that the scale of the vertical axis differs across plots. (Matrix Weibull Model)}
	\label{fig:evolution-Weibull}
\end{figure*}

Figure~\ref{fig:density-Weibull} provides a visual comparison between the theoretical density of the true matrix Weibull distribution (using the parameters given in \cite{BladtPeralta2025}) and the density obtained using the estimated parameters (from Table \ref{tab:estimators-Weibull}). The close agreement between the two curves confirms the accuracy of the estimation procedure. The estimated density not only tracks the mode and decay pattern of the true model, but also captures the multi-modal structure introduced by the inhomogeneity.

\begin{figure*}[htbp]
	\centering
	\includegraphics[width=0.75\textwidth]{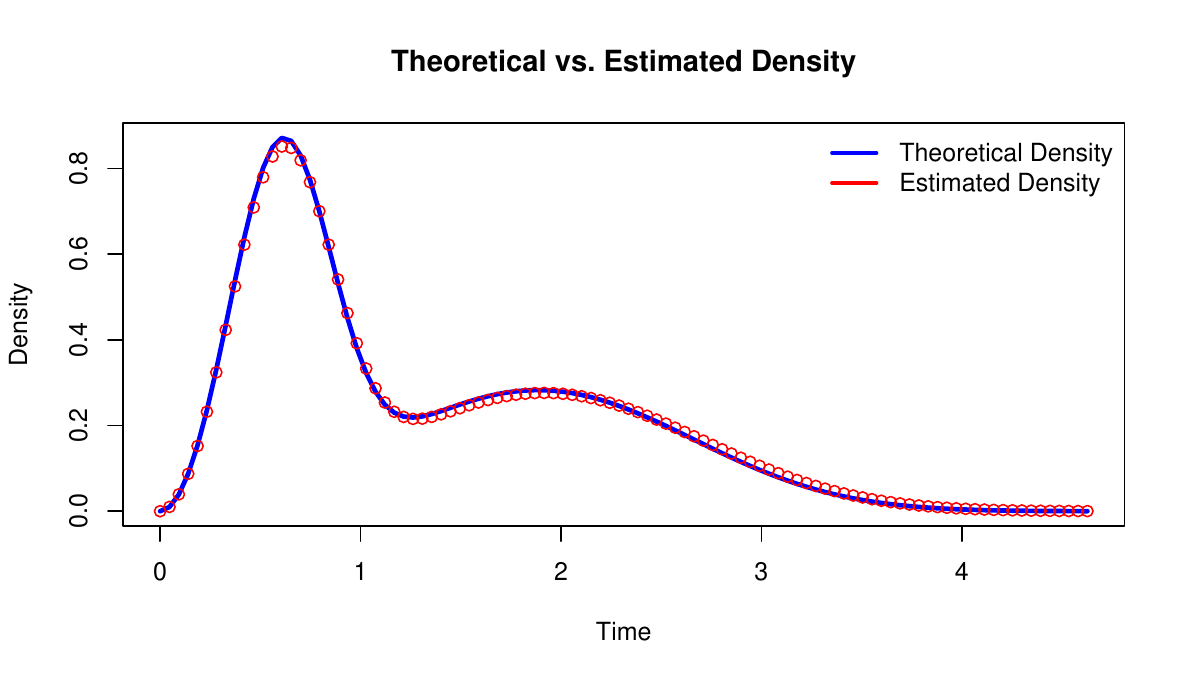}
	\caption{Comparison of the true theoretical density (blue) with the estimated density (red) for the matrix Weibull model.}
	\label{fig:density-Weibull}
\end{figure*}

\section{Real Data Application}\label{sec:real}
The dataset comes from monitoring a group of heart transplant recipients, available in the msm package in R \citep{msm}. It contains records for 622 patients, tracking the progression of coronary allograft vasculopathy (CAV), a post-transplant deterioration of the arterial walls. Each patient underwent periodic angiograms, typically once a year after transplantation, to diagnose and monitor CAV development. Based on the angiogram results, each patient's condition was classified into one of three possible states: CAV-free, mild CAV, or moderate/severe CAV. A fourth state, death, represents the absorbing state and is recorded at the time of the patient's death, measured in years since transplantation.

All patients begin in state 1 (CAV-free). Among the 622 records, only 251 patients reached the absorbing state (death). For model fitting, we considered the 251 absorbed records and the 15 non-absorbed records that transitioned through all three transient states: CAV-free, mild CAV, and moderate CAV. Records that did not visit these three states were excluded, as they provide limited information about the full disease-progression pathway considered in this illustrative analysis. The resulting dataset consists of discrete-time observations of each of the 266 patients' CAV status at the times of their medical visits. Importantly, the observation times are irregular: the intervals between visits, denoted by $t_{k,i} - t_{k,i-1}$, vary both across and within patients. This irregular sampling poses a challenge for parameter estimation, since the exact transition times between states are unobserved and differ across trajectories; precisely the type of setting for which our proposed SEM-based inference method is designed. We divided the dataset into training and test samples, with 132 and 134 patients, respectively.

A further feature of the dataset is that some patients share exactly the same recorded absorption time (i.e., time of death). We retain these repeated absorption times in our analysis, provided that the corresponding patients differ in their intermediate observed states and/or visit times. Even when two individuals die at the same recorded time, their intermediate trajectories may differ substantially in both the spacing of visits and the sequence of observed CAV states, and both features carry meaningful information for estimating transition intensities. Removing such observations would therefore discard informative patterns in the data. Moreover, because the train-test split is random, repeated absorption times naturally appear in both subsets, further supporting the decision to keep the dataset intact.

The choice of the scaling function $h_\beta(t) = e^{\beta t}$ is motivated by the well-established success of Gompertz-type dynamics in modeling human mortality and other biological aging processes. The Gompertz distribution has long been used to describe lifetimes and failure rates that increase exponentially with time \citep{gompertz1825}, and recent studies have shown that inhomogeneous phase-type distributions provide accurate representations of mortality curves across the entire lifespan \citep{Bladt2022}. Furthermore, Gompertz-type dynamics naturally arise in interacting failure systems, supporting their application to chronic vascular processes \citep{Nielsen2024}. This distribution has also been widely applied in survival analyses of cancer clinical trials with long-term survivors, illustrating its flexibility in modeling asymmetric hazard trajectories \citep{Tahira2024}. Collectively, these findings motivate the use of an exponential scaling function to capture the accelerating risk of CAV progression following transplantation.

We ran an adapted version of Algorithm \ref{main.algorithm} with the scaling function $h_{\beta}(t) = e^{\beta t}$, initial value $\hat{\beta}_0=2$, and initial sub-intensity matrix $\boldsymbol{\hat{\Lambda}}_0$ obtained by applying the maximum likelihood estimator for homogeneous MJPs described in Proposition~\ref{prop:MLEs}, treating the discrete-time observations $\bm{X}^d$ as if they were continuously observed paths from a homogeneous process (i.e., no time transformation was applied). Because this dataset contains exact absorption times rather than interval-censored ones\footnote{See Remark~\ref{rmk:incomplete.censored.data}}, we conditioned directly on these known exact times, bypassing the need to simulate them during the SE-step as described in the general algorithm. Note that since all patients start in state 1, the initial distribution is fixed as $\bm{\hat{\pi}} = \left(1, 0, 0\right) $. The learning rate was set to $\eta=10^{-5}$ and the convergence threshold to $ e_\ell = 10^{-4}$. The algorithm reached convergence after 21 iterations, yielding the estimates $\hat{\beta}=0.0848$ and
\[
\hat{\bm \Lambda}=
\begin{pmatrix}
	-0.1888 & 0.0871 & 0.0073 \\
	0.0962 & -0.3499 & 0.1749 \\
	0.0144 & 0.0288 & -0.1659
\end{pmatrix}.
\]

The estimated sub-intensity matrix $\boldsymbol{\hat{\Lambda}}$ characterizes the baseline transition structure among the three transient states describing CAV progression. Since the inhomogeneous model assumes $\Lambda(t) = e^{\beta t} \Lambda$, the instantaneous transition intensities increase exponentially with time. Thus, the entries of $\boldsymbol{\hat{\Lambda}}$ reflect the relative strength of transitions, while the overall progression of CAV accelerates with time since transplantation.

The largest off-diagonal elements correspond to transitions out of the mild CAV state: progression to moderate/severe CAV ($\hat{\lambda}_{23} = 0.1749$) and regression to the CAV-free state ($\hat{\lambda}_{21} = 0.0962$). The relatively high rate of regression from state 2 to state 1 is, according to \cite{msm}, a known characteristic of this dataset attributed to misclassification in angiogram readings. The initial onset of the disease from CAV-free to mild CAV occurs at a rate of $\hat{\lambda}_{12} = 0.0871$. Direct transitions between nonadjacent states are comparatively rare, particularly the jump from CAV-free directly to moderate/severe CAV ($\hat{\lambda}_{13} = 0.0073$).

Interestingly, state 2 (mild CAV) is the most transient one: it has the largest holding rate ($|\hat{\lambda}_{22}| = 0.3499$), indicating that patients spend the least time in state 2. In contrast, state 3 has the smallest total holding rate ($|\hat{\lambda}_{33}| = 0.1659 $), implying the longest expected duration in that state. This relative stability may reflect behavioral adaptation: once patients are diagnosed with moderate/severe CAV, they often receive closer monitoring and modify their lifestyle or treatment, slowing further progression.

The exit $\boldsymbol{\hat{\lambda}} = - \boldsymbol{\hat{\Lambda}} \mathbbm{1}_3$, which here represent the instantaneous rates of death from each transient state, are:
\[
\hat{\lambda}_1 = 0.0944, \quad 
\hat{\lambda}_2 = 0.0788, \quad 
\hat{\lambda}_3 = 0.1227.
\]
These values suggest that patients in the moderate/severe CAV state exhibit the highest instantaneous risk of death, followed by those in the CAV-free state. The relatively smaller absorption rate for the mild CAV state is consistent with the fact that most exits from this state are progressions to moderate/severe CAV rather than death.

To evaluate the accuracy of the estimated parameters and the adequacy of the chosen scaling function $h_{\beta}(t)$, we used the fitted model to generate simulated data and compared it with the real observations. Specifically, we simulated complete trajectories until absorption (all simulations reached the absorbing state) and compared the resulting absorption times with those observed in the test sample. Although the test sample contains 134 patients, for this particular comparison, we restricted the data to the 122 patients who both reached the absorbing state and had unique absorption times. Removing tied observations is a standard and necessary step to satisfy the continuity assumption of the two-sample Kolmogorov-Smirnov (KS) test, ensuring the validity of the computed p-value. The goodness-of-fit was assessed using the KS test, as in the examples in Section~\ref{Sect.Simulation.Study}. The resulting p-value of $0.5966$ indicates no statistically significant departure between the empirical distribution of absorption times generated under the estimated parameters and that observed in the real data. This behavior is further illustrated in Figure~\ref{fig:absorption-distributions-real-data}, which compares the empirical distributions of the observed absorption times with those obtained from the simulated data under the estimated model.

\begin{figure*}[htbp]
	\centering
	\includegraphics[width=0.75\textwidth]{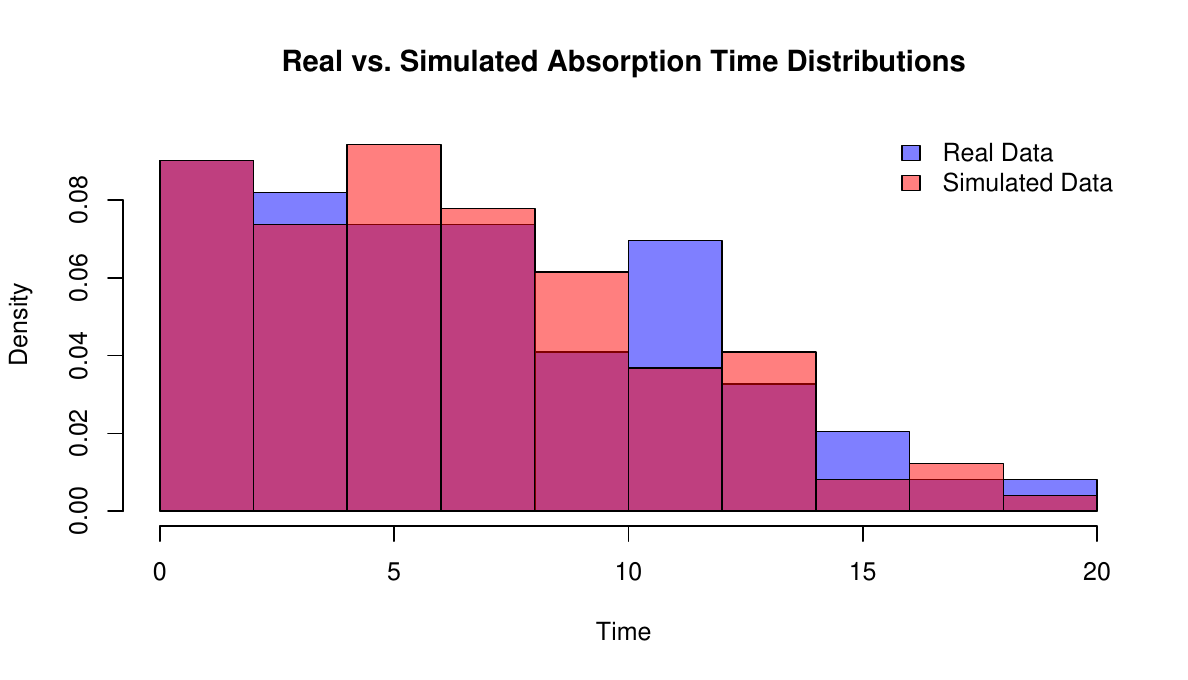}
	\caption{Comparison of the empirical distributions of absorption times for the real (test sample) and simulated data under the estimated parameters assuming the Gompertz model. The KS test yields a p-value of 0.5966.}
	\label{fig:absorption-distributions-real-data}
\end{figure*}

The results obtained from the real dataset confirm the practical relevance of the proposed estimation procedure. The inhomogeneous phase-type model, combined with the scaling function $h_{\beta}(t) = e^{\beta t}$, provides an accurate description of the progression of CAV over time and captures the increasing risk dynamics associated with disease advancement. Overall, the close agreement between simulated and observed absorption times provides evidence that the fitted model captures the marginal absorption-time behavior of the data. Additional path-level diagnostics could be used to further assess the fitted transient-state dynamics.

To assess the importance of accounting for time inhomogeneity, we also fitted a homogeneous phase-type model to the same dataset, assuming constant transition intensities over time. In this case, the sub-intensity matrix $\boldsymbol{\Lambda}(t)$ was replaced by a time-invariant sub-intensity matrix $\boldsymbol{\Lambda}$, which was estimated using a simplified version of Algorithm~\ref{main.algorithm}. Specifically, the estimation procedure omitted the parameter $\beta$ and the time transformations governed by the functions $g$ and $g^{-1}$, which were set to the identity. Since the convergence criterion in Algorithm~\ref{main.algorithm} depends on likelihood updates involving $\beta$ (see equation~\eqref{eq:GD.convergence.criteria}), there was no need to specify a learning rate $\eta$ or a convergence threshold $e_\ell$. Instead, we ran the algorithm for 300 iterations to allow sufficient warm-up and obtained the final estimate of $\boldsymbol{\Lambda}$ as the average of the last 20 iterations. This adaptation ensures comparability with the inhomogeneous case while preserving the stochastic structure of the SEM-based estimation procedure.

As in the inhomogeneous case, we used the fitted homogeneous model to generate simulated trajectories until absorption and compared them with the real observations (using the same restricted test sample of 122 unique absorption times as before to satisfy the KS test assumptions). As expected, the homogeneous model provided a noticeably poorer fit: the empirical and simulated absorption time distributions differed substantially, particularly in the tails, as shown in Figure~\ref{fig:absorption-distributions-real-data-homogeneous}. The corresponding KS test yielded a significantly lower p-value of 0.01066, leading to the rejection of the null hypothesis that the two samples come from the same distribution.

\begin{figure*}[htbp]
	\centering
	\includegraphics[width=0.75\textwidth]{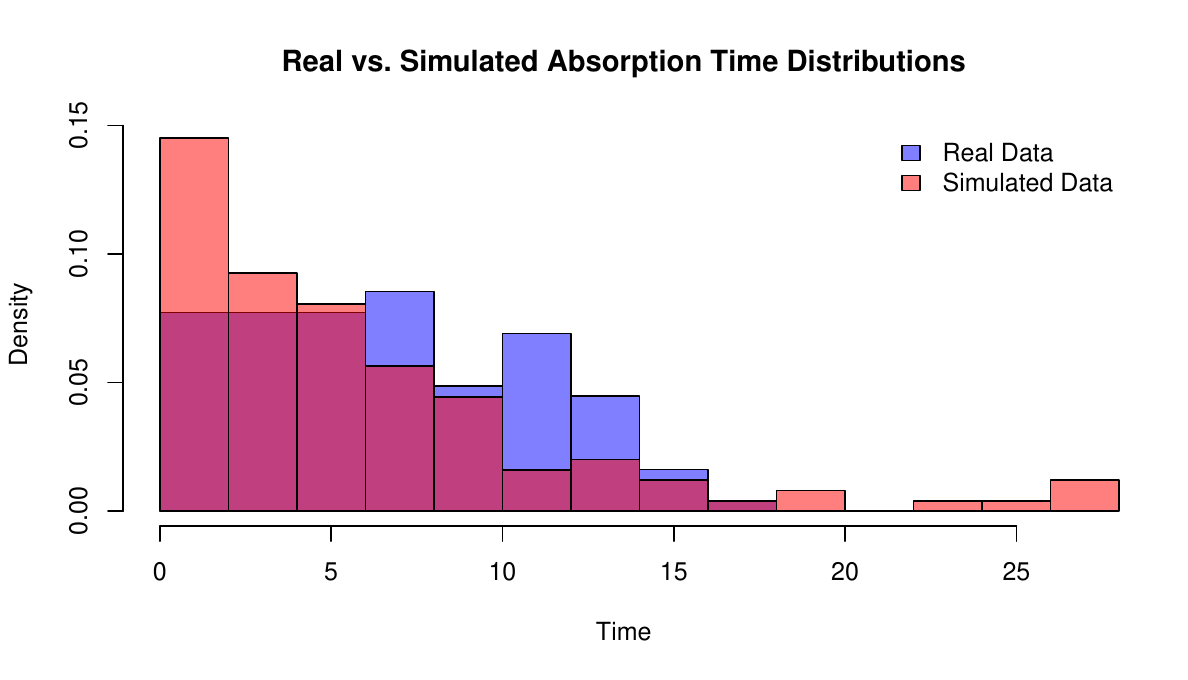}
	\caption{Comparison of the empirical distributions of absorption times for the real (test sample) and simulated data under the estimated parameters assuming homogeneity. The KS test yields a p-value of 0.01066.}
	\label{fig:absorption-distributions-real-data-homogeneous}
\end{figure*}

This comparative analysis highlights that our SEM-based inference method not only captures the temporal evolution of transition intensities but also yields reliable parameter estimates even under irregular and discretely observed data, establishing a robust foundation for future applications of inhomogeneous phase-type models.

\backmatter

%
%
%

\bmhead{Acknowledgements}
The authors would like to thank Zhihao Zhao for his careful reading of an earlier version of this paper and for his valuable suggestions.

\section*{Declarations}

%
%

\bmhead{Competing interests}
The manuscript has not been published elsewhere, nor is it under consideration by any other journal. All authors have approved the manuscript, and we have no conflicts of interest to disclose regarding this submission.

\bmhead{Funding}
Fernando Baltazar-Larios was supported by UNAM-DGAPA-PAPIIT-IN110926.

Alejandra Quintos was supported by the Office of the Vice Chancellor for Research and Graduate Education at the University of Wisconsin-Madison with funding from the Wisconsin Alumni Research Foundation.

\bmhead{Data availability}
All codes underlying the findings in our manuscript are publicly available through a \href{https://github.com/alequintos/Parameter_Estimation_for_IPH}{Github repository}
\begin{center}
    \url{https://github.com/alequintos/Parameter_Estimation_for_IPH}
\end{center}
The repository includes the R scripts used for simulation, inference, and figure generation
. The real dataset used in this study is publicly available, and its source is cited within the manuscript.

\noindent
%
%
%
%
%

\bibliography{sn-bibliography}

\end{document}